
\mag=1200


\font\seventeenbf=cmbx12 at 14pt

\font\tenrm=cmr10
\font\tenit=cmti10
\font\tensl=cmsl10
\font\tenbf=cmbx10
\font\tentt=cmtt10
\font\Bbb=msbm10
\font\Aaa=msam10

\def\tenpoint{%
\def\rm{\fam0\tenrm}%
\def\it{\fam\itfam\tenit}%
\def\sl{\fam\slfam\tensl}%
\def\tt{\fam\ttfam\tentt}%
\def\bf{\fam\bffam\tenbf}%
}

\tenpoint\rm

\newcount\notenumber

\def\note{\footnote{{\mathsurround=0pt}}}

\normalbaselineskip=11.66pt
\normallineskip=2pt minus 1pt
\normallineskiplimit=1pt
\normalbaselines
\vsize=17.3cm
\hsize=12.1cm
\parindent=20pt
\smallskipamount=3.6pt plus 1pt minus 9pt
\abovedisplayskip=1\normalbaselineskip plus 3pt minus 9pt
\belowdisplayskip=1\normalbaselineskip plus 3pt minus 9pt
\skip\footins=2\baselineskip
\advance\skip\footins by 3pt

\mathsurround 2pt
\newdimen\leftind \leftind=0cm
\newdimen\rightind \rightind=0.65cm

\def\pagenumbers{\footline={\hss\tenrm\folio\hss}}
\nopagenumbers

\def\MainHead{{\baselineskip 16.7pt\seventeenbf
\noindent\titolo\par}
\normalbaselines
\vskip 18.34pt
\noindent\autori\par
\ni\indirizzo
\footnote{\phantom{i}}{\piedipagina}
\vskip 23.34pt
\ni {\bf Abstract.} \Abstract
}

%
%
%
%
%
\font\ChapTitle=cmbx12
\font\SecTitle=cmbx12
\font\SubSecTitle=cmbx12
%
%
%
\def\NormalSkip{\parskip=5pt\parindent=15pt\baselineskip=12pt\PageNumbers%
\leftskip=0cm\rightskip=0cm}

\def\PageNumbers{\footline={\hss\tenrm\folio\hss}}
\def\NoPageNumbers{\footline={}}

%
%
%
\def\np{\vfill\eject}
\def\ss{\vskip 5pt}
\def\ms{\vskip 15pt}
\def\bs{\vskip 30pt}

\def\ni{\noindent}
%
%
%
\newcount\CHAPTER		          %
\newcount\SECTION		          %
\newcount\SUBSECTION		       %
\newcount\FNUMBER     
%
%
%
\CHAPTER=0		          
\SECTION=0		          
\SUBSECTION=0		       
\FNUMBER=0		          
%
%
%
\long\def\NewChapter#1{\global\advance\CHAPTER by 1%
\global\edef\CurrentChapter{\the\CHAPTER}%
\np\NoPageNumbers\ \vfil\parindent=0cm\leftskip=2cm\rightskip=2cm{\ChapTitle #1\hfil}%
\vskip 4cm\ \vfil\eject\SECTION=0\SUBSECTION=0\FNUMBER=0\NormalSkip}
%
%
%

\long\def\NewSection#1{\bs\global\advance\SECTION by 1%
\global\edef\CurrentSection{\the\SECTION}%
\ni{\SecTitle \ifnum\CHAPTER>0 \CurrentChapter.\fi\CurrentSection.\ #1}\ms\SUBSECTION=0\FNUMBER=0}
%
%
\long\def\NewSubSection#1{\global\advance\SUBSECTION by 1%
\global\edef\CurrentSubSection{\the\SUBSECTION}%
{\SubSecTitle \ifnum\CHAPTER>0
\CurrentChapter.\fi\CurrentSection.\CurrentSubSection.\ #1}\ss\FNUMBER=0}

%
%
\def\HeadNumber{\ifnum\CHAPTER>0 \CurrentChapter.\fi%
\ifnum\SECTION>0 \CurrentSection.\ifnum\SUBSECTION>0 \CurrentSubSection.\fi\fi}
\def\ComposedRightFormulaNumber{\global\advance\FNUMBER by 1%
\eqno{\fopen\HeadNumber\the\FNUMBER\fclose}}
\def\ComposedLeftFormulaNumber{\global\advance\FNUMBER by 1%
\leqno{\fopen\HeadNumber\the\FNUMBER\fclose}}
\def\ComposedRightFormulaLabel#1{\global\advance\FNUMBER by 1%
\eqno{\fopen\HeadNumber\the\FNUMBER\fclose}%
\global\edef#1{\fopen\HeadNumber\the\FNUMBER\fclose}}
\def\ComposedLeftFormulaLabel#1{\global\advance\FNUMBER by 1%
\leqno{\fopen\HeadNumber\the\FNUMBER\fclose}%
\global\edef#1{\fopen\the\FNUMBER\fclose}}

%
%
\def\ComposedTheoremNumber{\global\advance\FNUMBER by 1 \fopen\HeadNumber\the\FNUMBER\fclose}
\def\ComposedTheoremLabel#1{\global\advance\FNUMBER by 1\fopen\HeadNumber\the\FNUMBER\fclose%
\global\edef#1{\fopen\HeadNumber\the\FNUMBER\fclose}}
%
%
%
\def\fopen{(}\def\fclose{)}                 
\def\fn{\ComposedRightFormulaNumber}        
\def\fl{\ComposedRightFormulaLabel}         

\def\Compare#1#2{\message{^^J Compara \noexpand #1:=#1[#2]^^J}}

%
%
%
%

\def\ni{\noindent}
\def\ss{\vskip 5pt}
\def\ms{\vskip 10pt}

\def\noex{\noexpand}

\def\refs{}
\def\empty{\#}
\def\BibNumber{}
\def\BibTitle{}

\newcount\BNUM
\BNUM=0

\def\bib#1#2{\gdef#1{\global\def\BibNumber{\empty}\global\def\BibTitle{#2}}}

\def\ref#1{#1
\if\BibNumber\empty \global\advance\BNUM 1
\message{reference[\BibNumber]}\message{}
\global\edef\refs{\refs \ss\ni[\the\BNUM]\ \BibTitle}
\global\edef#1{\noex\global\noex\edef\noex\BibNumber{[\the\BNUM]}
 \noex\global\noex\edef\noex\BibTitle{\BibTitle}}
{\bf [\the\BNUM]}
\else
{\bf \BibNumber}
\fi}

\def\Biblio{{\refs}}



\def\Lor{\hbox{\rm Lor}}

\def\ds{\hbox{\bf ds}}
\def\d{\hbox{\rm d}}
\def\dt{\hbox{\rm dt}}
\def\dr{\hbox{\rm dr}}

\def\dphi{\hbox{\rm d$\phi$}}

\def\dim{\hbox{\rm dim}}
\def\det{\hbox{\rm det}}

\def\fraz#1#2{{\textstyle {{#1}\over{#2}}}}

\def\Lie{\hbox{\it \$}}

\def\calC{{\cal C}} 
\def\calE{{\cal E}}

\def\calL{{\cal L}} 
\def\calW{{\cal W}} 
 
\def\calU{{\cal U}} 
\def\calB{{\cal B}} 
\def\calQ{{\cal Q}}

\def\calK{{\cal K}}


\def\na{\nabla}
\def\la{\lambda}
\def\Si{\Sigma}
\def\si{\sigma}
\def\ka{\kappa}

\def\ep{\epsilon}
\def\al{\alpha}
\def\be{\beta}
\def\Ga{\Gamma}
\def\ga{\gamma}

\def\de{\delta}
\def\ze{\zeta}
\def\te{\theta}
\def\Te{\Theta}

\def\R{\hbox{\Bbb R}}

\def\one{\hbox{\Bbb I}}
\def\inner{\hbox{\Aaa \char121}}


\def\R{I \kern-.36em R}
\def\E{I \kern-.36em E}
\def\F{I \kern-.36em F}
\def\Co{I \kern-.66em C}
\def\id{1 \kern-.36em I}              

\def\del{\partial}                   

\def\QDE{{\offinterlineskip\lower1pt\hbox{\kern2pt\vrule width0.8pt
\vbox to8pt{\hbox to6pt{\leaders\hrule height0.8pt\hfill}\vfill%
\hbox to6pt{\hrulefill}}\vrule\kern3pt}}}


\def\arr{\rightarrow }            
\def\then{\quad\Rightarrow\quad}      
\def\QDE{\hbox{\ }\vrule height4pt width4pt depth0pt}                                                              

\def\np{\vfill\eject}
\def\ni{\noindent}

\def\ss{\vskip 5pt}
\def\ms{\vskip 10pt}
\def\bs{\vskip 15pt}

\bib{\Einstein}  
{A. Einstein, 
{\it Sitzungsber. Preuss. Akad. Wiss. (Berlin)},
1111, (1916).}

\bib{\TaubNUTH}{C.\
J.\ Hunter, hep-th/9807010}

\bib{\TaubNUTHH}{S.\ W.\ Hawking, C.\ J.\ Hunter, hep-th/9808085}

\bib{\TaubNUTHHP}{ S.\ W.\ Hawking, C.\ J.\ Hunter, D.\ N.\ Page,
hep-th/9809035}

\bib{\HawHun}{ S.\ W.\ Hawking, C.\ J.\ Hunter, Class. Quantum Grav. {\bf 13}, (1996) 2735}

\bib{\HawHor}{ S.\ W.\ Hawking, G.\ T.\ Horowitz, Class. Quantum Grav. {\bf 13}, (1996)
1487}

\bib{\TaubNUTP}{ D.\ N.\ Page, Phys.\ Lett.\ {\bf 78B}, (1978) 249}

\bib{\Misner}{ C.\ W.\ Misner, J.\ Math.\ Phys.\ {\bf 4}, (1963) 924}

\bib{\WaldA}{V. Iyer and R. Wald, Phys. Rev. D {\bf 50},  (1994) 846;
R.M.\ Wald, J.\ Math.\ Phys., {\bf 31}, (1993) 2378}

\bib{\Trautman} {A. Trautman, in: {\it Gravitation: An Introduction to Current Research}, L.
Witten ed.
(Wiley, New York, 1962) 168}

\bib{\Trautmandue} {A. Trautman,
  Commun. Math. Phys., {\bf  6}, (1967)    248}

\bib{\Komar} {A.  Komar, Phys. Rev., {\bf  113}, 
934, (1959).}

\bib{\WaldB}{I.\ Racz, R.\ M.\ Wald, Class. Quantum Grav. {\bf 9}, (1992) 2643}

\bib{\Remarks}{
L.\ Fatibene, M.\ Ferraris, M.\ Francaviglia, M.\ Raiteri,  Annals of Phys., 
{\bf 275}, (1999) 27
}

\bib{\BTZ}{
L.\ Fatibene, M.\ Ferraris, M.\ Francaviglia, M.\ Raiteri,  Phys. Rev. D {\bf 60},
(1999) 124012}

\bib{\BCEA}{
L.\ Fatibene, M.\ Ferraris, M.\ Francaviglia, M.\ Raiteri,  Phys. Rev. D {\bf 60},
(1999) 124013}

\bib{\TaubBolt}{ L.  Fatibene, M.  Ferraris, M.  Francaviglia, M.  Raiteri,
gr-qc/$9906114$, Annals of Phys. (in press)}

\bib{\Kolar}{I.\ Kol{\'a}{\v r}, P.\ W.\ Michor, J.\ Slov{\'a}k, 
{\it Natural Operations in Differential Geometry}, 
Springer--Verlag, (New York, 1993)
}

\bib{\Saunder}{D.J.\ Saunders, {\it The Geometry of Jet Bundles},
Cambridge University Press, (Cambridge, 1989)
}

\bib{\Lagrange}{M. Ferraris, M. Francaviglia,
in: {\it Mechanics, Analysis and Geometry: 200 Years after Lagrange},
Editor: M. Francaviglia, Elsevier Science Publishers B.V., (Amsterdam, 1991) 451}

\bib{\Cavalese}{M. Ferraris and M. Francaviglia, in: {\it 8th Italian
Conference on General Relativity and Gravitational Physics}, Cavalese (Trento), August 30 --
September 3, World Scientific, (Singapore, 1988) 183; M. Ferraris and M. Francaviglia, Gen. Rel.
Grav.  {\bf 22}
(9), (1990) 965}

\bib{\Palatini}{M. Ferraris, M. Francaviglia, C. Reina, Gen. Rel.
Grav.  {\bf 14}, (1982) 243}

\bib{\Ferraris}{M. Ferraris, in:
{\it Proceedings of the Conference on Differential Geometry and Its Applications},
Part 2, Geometrical Methods in Physics,
D.\ Krupka ed., (Brno, 1984) 61}

\bib{\Robutti}{M.\ Ferraris, M.\ Francaviglia and O.\ Robutti, in:{\it G\'eom\'etrie et Physique},
Proceedings of the {\it Journ\'ees Relativistes 1985} (Marseille, 1985), 
Y.\ Choquet--Bruhat, B.\ Coll, R.\ Kerner, A.\ Lichnerowicz eds. Hermann, (Paris, 1987) 112}

\bib{\Katz}{J.\ Katz, Class.\ Quantum Grav., {\bf 2}, (1985) 423}

\bib{\KatzBicak}{J.\ Katz, J.\ Bicak, D.\ Lynden--Bell,  Phys.\ Rev.\ D{\bf 55}
(10), (1997) 5957}

\bib{\KatzBondi}{J.\ Katz, D.\ Lerer, gr-qc/$9612025$}

\bib{\PetrovKatz}{A.\ N.\ Petrov, J.\ Katz,  gr-qc/$9905088$}

\bib{\CADM}{M.\ Ferraris and M.\ Francaviglia, Atti Sem. Mat. Univ. Modena, {\bf 37},
1989, 61} 

\bib{\Sinicco}{M.\ Ferraris, M.\ Francaviglia and I.\ Sinicco, Il Nuovo Cimento, {\bf 107B},
(11), 1992, 1303}

\bib{\CADMC}{M.\ Ferraris and M.\ Francaviglia, Gen.\ Rel.\ Grav., {\bf 22}, (9), (1990) 965}

\bib{\York}{ J.\ W.\ York, Foundation of Phys., {\bf 16} (3), (1986) 249}

\bib{\Hayward}{G.\ Hayward, K.\ Wong,  Phys.\ Rev.\ D{\bf 46} (2), (1992) 620}

\bib{\Haywardnon}{G.\ Hayward,  Phys.\ Rev.\ D{\bf 47} (8), (1993) 3275}

\bib{\Mannultimo}{I.\ S.\ Booth, R.\ B.\ Mann, gr-qc/$9907072$}

\bib{\Nester}{C.\ M.\ Chen, J.\ M.\ Nester,  gr-qc/$9809020$}

\bib{\BY}{J.\ D.\ Brown, J.\ W.\ York, Phys.\ Rev.\ D{\bf 47} (4), (1993) 1407}

\bib{\BYdue}{J.\ D.\ Brown, J.\ W.\ York, Phys.\ Rev.\ D{\bf 47} (4), (1993) 1420}

\bib{\GH}{G.\ W.\ Gibbons, S.\ W.\ Hawking, Phys.\ Rev.\ D{\bf 15} (10), (1977) 2752}

\bib{\Brown}{ J.\  D.\  Brown, gr-qc/9506085.}

\bib{\RT}{T.\ Regge, C.\ Teitelboim, Annals of Physics {\bf 88}, (1974) 286.}
 
\bib{\ADM} {R.  Arnowitt, S.  Deser and C.  W.  Misner, in: 
{\it Gravitation: An Introduction
to Current Research}, L. Witten ed. Weley,  227, (New York, 1962).}

\bib{\II} {J.\ Isenberg, J.\ Nester, in: 
{\it General Relativity and Gravitation. One Hundred Years after 
the Birth of Albert Einstein},  Vol.{\bf 1}, A. Held ed., Plenum 
Press, 23, (New York 1980).}

\bib{\Gravitation} {C.W. Misner, K.S. Thorne, J.A. Wheeler, 
{\it Gravitation}, (Freeman, San Francisco,  1973)}

\bib{\BTZRefB}{J.\ D.\ Brown, J.\ Creighton, R.\ B.\ Mann, Phys. Rev. D{\bf 50}, 6394 (1994)}

\bib{\AltriTaubMann}{
R.\ B.\ Mann, hep-th/9903229; R.\ B.\ Mann, hep-th/9904148.}

\bib{\Waldqc}{V. Iyer and R. Wald
gr-qc/9503052.}

\bib{\McLenaghan}{
K.\  Chu, C.\ Farel, G.\ Fee, R.\ McLenaghan, Fields Inst. Comm. {\bf 15}, (1997) 195
}

%
%
%
\def\pspicture #1 by #2 (#3){
  \vbox to #2{
    \hrule width #1 height 0pt depth 0pt
    \vfill
    \includegraphics{#3} 
    }
  }
\def\scaledpspicture #1 by #2 (#3 scaled #4){{
  \dimen0=#1 \dimen1=#2
  \divide\dimen0 by 1000 \multiply\dimen0 by #4
  \divide\dimen1 by 1000 \multiply\dimen1 by #4
  \pspicture \dimen0 by \dimen1 (#3)}
  }
%
%
%
\def\Tubo{\scaledpspicture 358pt by 508pt  (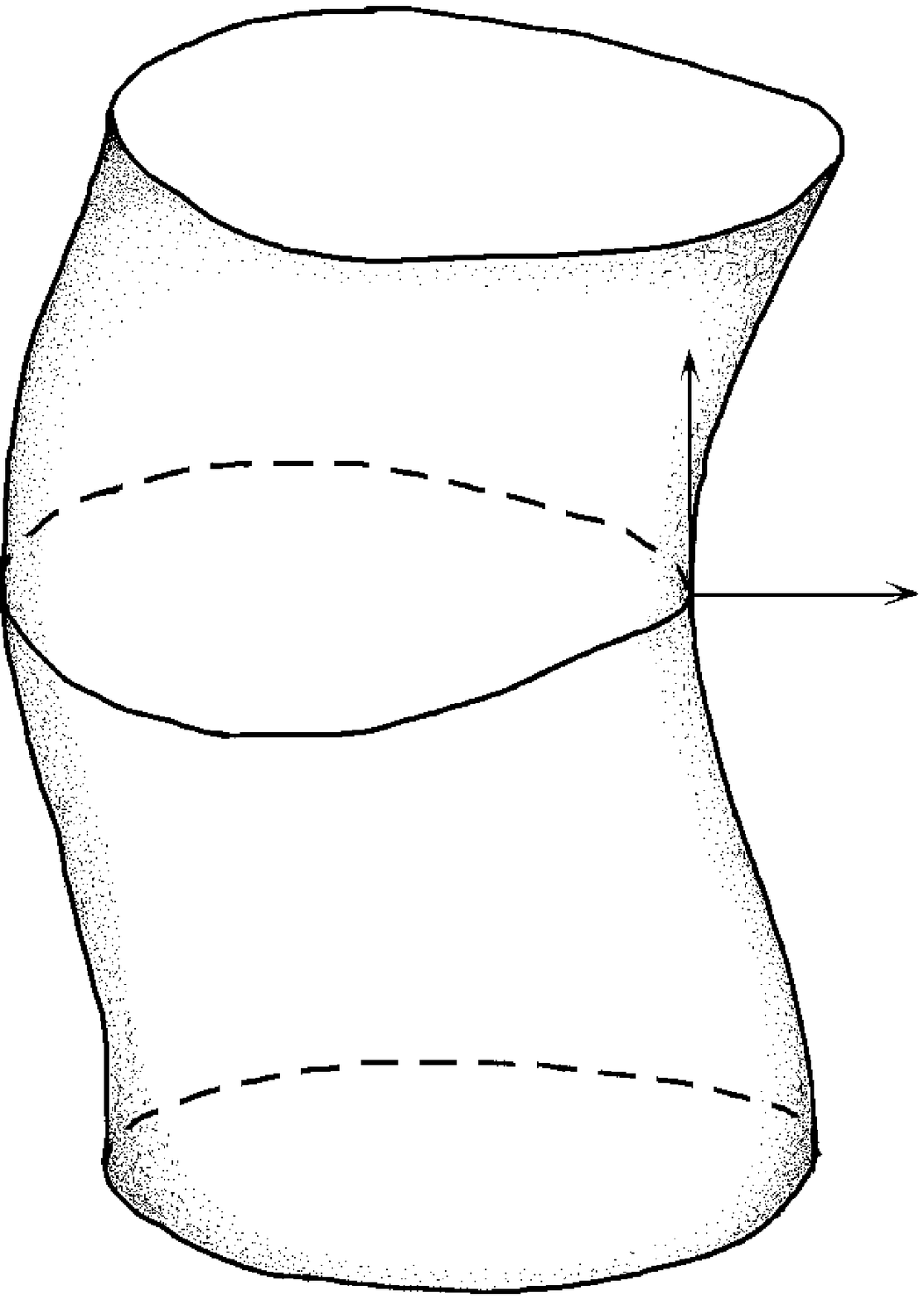 hscale=30 vscale=30 scaled 300)}
\def\TuboDritto{\scaledpspicture 320pt by 451pt  (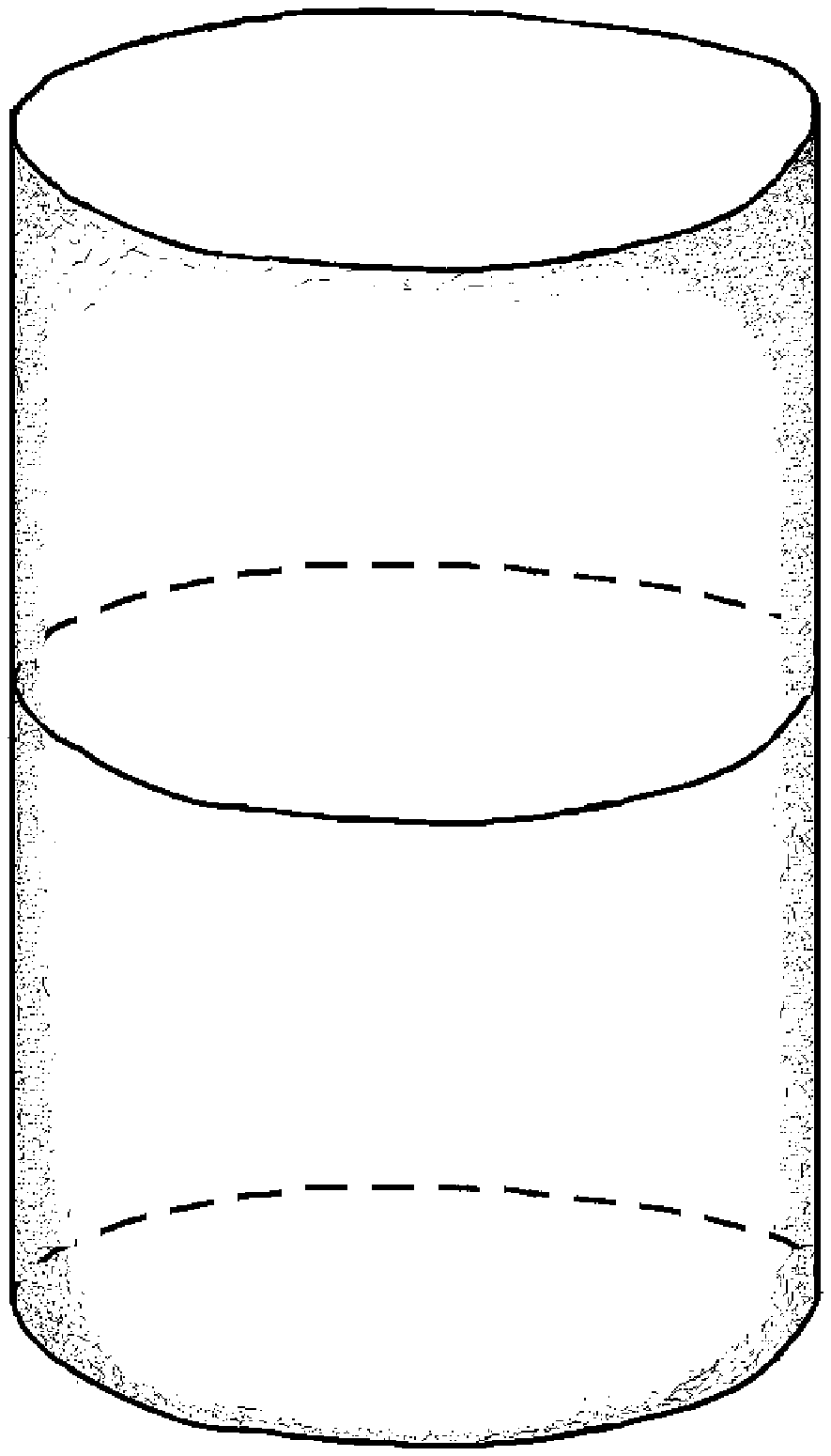 hscale=25 vscale=25 scaled 258)}
\def\TuboStorto{\scaledpspicture 331pt by 470pt  (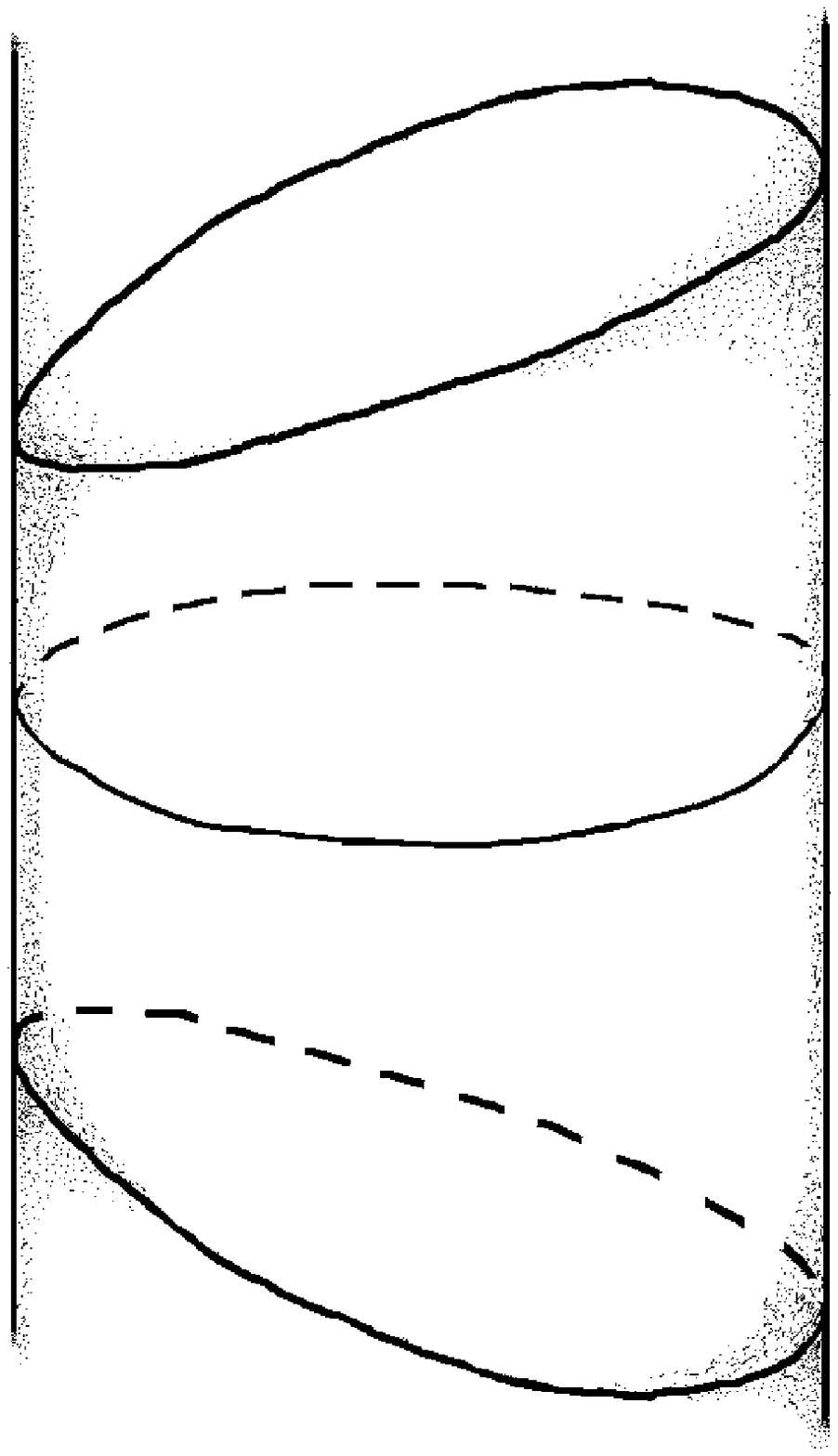 hscale=25 vscale=25 scaled 250)}
%


\def\titolo{N\"other Charges, Brown--York Quasilocal
Energy and Related Topics}
\def\autori{%
L.\ FATIBENE
\note{E-mail: fatibene@dm.unito.it, E-mail: ferraris@dm.unito.it,},
M.\ FERRARIS,
M.\ FRANCAVIGLIA\note{E-mail: francaviglia@dm.unito.it, E-mail: raiteri@dm.unito.it},
M.\ RAITERI}

\def\indirizzo{Dipartimento di Matematica, Universit\`a degli Studi di Torino,\par
\ni Via Carlo Alberto 10, 10123 Torino, Italy}

\def\Abstract{
The Lagrangian proposed by York et al.\ and the covariant first order Lagrangian for General
Relativity are introduced to deal with the (vacuum) gravitational field on a reference background.
The two Lagrangians are compared and we show
 that the first one can be obtained from the latter under suitable hypotheses.
The induced variational principles are also compared and discussed.
A conditioned correspondence among N\"other conserved quantities, quasilocal energy and the
standard Hamiltonian obtained by $3+1$ decomposition is also established.
As a result, it turns out that the covariant first order Lagrangian is better suited whenever a
reference background field has to be taken into account, as it is commonly accepted when  dealing
with conserved quantities in non--asymptotically flat spacetimes. As a further advantage of
the use of a covariant first order Lagrangian, we show that all the  quantities computed are
manifestly covariant, as it is appropriate in General Relativity. }

\def\piedipagina{}

\vglue 63.3pt
\raggedbottom

\MainHead

\pagenumbers

\NewSection{Introduction}

Many approaches to variational principles, conserved quantities and related topics  can be found in
the current literature about General Relativity (see, e.g., \ref{\Trautman}, \ref{\Lagrange},
\ref{\BY}, \ref{\ADM},
\ref{\RT}, \ref{\Gravitation}, \ref{\York}, \ref{\Haywardnon},  \ref{\Mannultimo},
\ref{\HawHun}, \ref{\KatzBondi},  \ref{\KatzBicak}, \ref{\PetrovKatz}\ and references quoted
therein). We shall hereafter compare two of them, both dealing with a dynamical metric
$g$ and a reference background metric
$\bar g$  over a spacetime manifold $M$ of  arbitrary dimension $n$, with $n\ge 3$. A large part of
the paper will deal with the explicit  case $n=4$, in view of applications  to standard General
Relativity. We shall deal with the vacuum case; the generalization to the case of presence of matter
fields is straightforward.

The first variational principle  we shall deal with is based on the covariant first order
action functional for General Relativity (see \ref{\KatzBicak}, \ref{\Katz}, \ref{\Cavalese},
\ref{\BTZ}). The second one ( \ref{\BY}, \ref{\York}) is based on the action functional due to
York et al., which is defined with the aim of dealing with the fixing of the 
$(n-1)$--metric induced on the boundary
$\del D$ of any region $D$ of  spacetime $M$.

The covariant first order action was introduced  to set General Relativity in
a standard  {\it covariant} first order variational framework. As  is well known, in fact,
the Hilbert action functional $L=(1/2\ka)\, \sqrt{g} \,R\, \ds$ is second order in the metric
field, so that field equations are expected to be of the fourth order. Einstein field equations
are second order equations instead, as if the action were  first order only. This is due to the
well known fact that {\it locally} second derivatives of the metric field appearing in the scalar
curvature may be hidden under a divergence, thus not appearing in field equations (a fact which
was clear to Einstein from the very beginning; see, e.g.  \ref{\Einstein}). Of
course, however, this cannot be done in general in a global and covariant way; that is why
General Relativity is usually considered as a second order field theory or, whenever it is
treated on a first order basis, something  is lost (e.g. covariance or boundary terms; see
\ref{\ADM}, \ref{\RT}).

The {\it covariant first order action functional} is the following:
$$
A_D[g,\bar g]= {1\over 2\ka}\int_D\sqrt{g}\> R\> \ds
-{1\over 2\ka}\int_{\del D}\sqrt{g} g^{\mu\nu} w^\al_{\mu\nu}\> \ds_\al
-{1\over 2\ka}\int_D\sqrt{\bar g}\> \bar R\> \ds
\fl{\FOL}
$$
where $\ka$ is a constant ($\ka=8\pi G/c^4$ in General Relativity with $\dim (M) =4$), $\sqrt{g}$ is
the square root of the absolute value of the determinant of the dynamical metric $g$,  $\sqrt{\bar
g}=\sqrt{\vert\det
\bar g\vert}$  the analogous quantity for the background metric $\bar g$ and
$\ds$ and
$\ds_\al=\del_\al\inner \ds$ are the standard local bases for $n$--forms and  $(n-1)$--forms over
$M$, respectively. We systematically denote by a bar the quantities referred to the background,
i.e.\ here and hereafter we shall use the following notation
$$
\matrix{
g_{\mu\nu}\hfill       &     \bar g_{\mu\nu}\hfill         &\hbox{covariant metric }\hfill    \cr
g^{\mu\nu}\hfill       &     \bar g^{\mu\nu}\hfill         &\hbox{contravariant metric }\hfill\cr
\Ga^\al_{\be\nu}\hfill  &\bar\Ga^\al_{\be\nu}\hfill &\hbox{Levi--Civita connection}\hfill \cr
R^\al_{\be\mu\nu}\hfill  &\bar R^\al_{\be\mu\nu}\hfill &\hbox{Riemann tensor }\hfill \cr
R_{\mu\nu}=R^\al_{\mu\al\nu} \hfill &\bar R_{\mu\nu}=\bar R^\al_{\mu\al\nu}\hfill &\hbox{Ricci
tensor}\hfill \cr R=g^{\mu\nu} R_{\mu\nu} \hfill &\bar R=\bar g^{\mu\nu}\bar R_{\mu\nu}\hfill
&\hbox{scalar curvature}\hfill \cr u^\mu_{\al\be}=
\Ga^{\mu}_{\al\be}-\de^\mu_{(\al}\Ga^{\ep}_{\be)\ep}\quad\hfill &
\bar u^\mu_{\al\be}= \bar \Ga^{\mu}_{\al\be}-\de^\mu_{(\al}\bar \Ga^{\ep}_{\be)\ep}\hfill &\cr
}
$$ 
Let us also introduce the following {\it relative quantities}:
$$
q^\mu_{\al\be}=\Ga^{\mu}_{\al\be}-\bar \Ga^{\mu}_{\al\be}\>,
\qquad\qquad
w^\mu_{\al\be}=u^{\mu}_{\al\be}-\bar u^{\mu}_{\al\be}
$$

The action functional $\FOL$ is associated to the so--called {\it covariant first order
Lagrangian}
\global\advance\FNUMBER by 1
\global\edef\FOELaguno{\fopen\HeadNumber\the\FNUMBER.a\fclose}
$$
L={1\over 2\ka}\Big(\sqrt{g}\> R\> 
-\d_\al\big(\sqrt{g} g^{\mu\nu} w^\al_{\mu\nu}\big)
-\sqrt{\bar g}\> \bar R\Big)\> \ds 
\eqno{\FOELaguno}
$$
or equivalently
\global\edef\FOELagdue{\fopen\HeadNumber\the\FNUMBER.b\fclose}
$$
L= {1\over2\ka}\Big((\sqrt{g}-\sqrt{\bar g})\bar R+
\sqrt{g} g^{\al\be}(q^\rho_{\al\si}q^\si_{\rho\be}-q^\si_{\si\rho}q^\rho_{\al\be})\Big)\>\ds
\eqno{\FOELagdue}
$$
From the first  expression $\FOELaguno$, it can be easily seen that the fields $g$ and $\bar g$ do
not interact and they both obey vacuum Einstein field equations (provided that suitable boundary
conditions are satisfied, as we shall discuss  shortly after).
From the second expression $\FOELagdue$, the Lagrangian is however recognised to be first order in
$g$ and second order in $\bar g$.
Being both $q^\mu_{\al\be}$ and $w^\mu_{\al\be}$ tensors, $L$ is a covariant Lagrangian.
It is, of course,  the truly covariant  counterpart of the so--called  Hilbert--Palatini first
order  Lagrangian (see \ref{\Palatini}), to which it reduces by suitable  non--covariant
cancellations and background fixings. We stress that in the variational principle induced by
$\FOL$ the dynamical metric
$g$ is endowed with a direct physical meaning, while the reference background metric $\bar g$ is,
at least for the moment, introduced to provide covariance and as a reference value for conserved
quantities, as discussed below. Notice that if the action is computed for $g=\bar g$, then it
identically vanishes.

We remark that the Hilbert--Einstein Lagrangian $L=(1/2\ka)\, \sqrt{g} \,R\, \ds$ is  a second order
Lagrangian and thus it would  {\it a priori} define a variational principle in which the
dynamical metric $g$ together with  its first derivatives are kept fixed on the boundary.
In fact, in standard variational calculus, for a covariant field theory described by a Lagrangian
of order $k$, one keeps fixed the fields  on the boundary  together with their derivatives up to
order
$k-1$ (see Appendix B).
However, the degeneracy of the Hilbert--Einstein action functional has been  globally  removed by
introducing a reference background metric $\bar g$ which provides a covariant way to cancel out the
second order derivatives of the dynamical metric $g$, as shown  in
$\FOELagdue$ above.
As we already remarked, the Lagrangian $\FOELagdue$ is first order in $g$ and
second order in $\bar g$ so that it defines a standard variational principle in which the
dynamical metric $g$ is kept fixed on the boundary, while the reference background
metric $\bar g$ is kept fixed together with its first derivatives on the boundary.
Of course, nothing more can be done without breaking down covariance, i.e.\ without
choosing an ADM foliation, or without requiring suitable additional matching conditions between the
two metrics.

On the other hand,  York's action  was originally built (see \ref{\BY}, \ref{\York},
\ref{\Hayward}) to provide a variational principle suited  to deal with the boundary
conditions which are specified by keeping the induced metric on
$\partial D$ fixed. Boundary terms are added to the standard Hilbert--Einstein action
$(1/2\ka)\,\int_D R\sqrt{g}\>\ds$. They  are exactly needed so that the boundary contribution to
the variation of the action vanishes when the induced $(n-1)$--metric is kept fixed on $\del D$.

The York's action functional  is adapted to an ADM foliation
induced by dragging a spacelike hypersurface $\Si$ along a timelike vector field $\ze$.  
{\bs
\ni\hskip .3truecm\Tubo
\ms
\ \hskip 1truecm {\sl Fig.\ 1}
\vskip-4.4truecm\hskip 3.8truecm $\vec n$
\vskip-1.7truecm\hskip 2.6truecm $\vec u$
\vskip 2.0truecm\hskip 3.9truecm $\calB=\cup_{_t} B_t$
\vskip 0.5truecm \hskip -.5truecm $B_{t_0}$\hskip 1.0truecm $\Si_{t_0}$
\vskip -3.3truecm \hskip -1.0truecm $B_{t}$\hskip 1.0truecm $\Si_{t}$
\vskip -2.8truecm \hskip -0.5truecm $B_{t_1}$\hskip 1.0truecm $\Si_{t_1}$
}
\vskip -1.84truecm \ \leftskip 7.5truecm

\ni
An ADM foliation of a spacetime region $D$
is obtained and it is pa\-ra\-me\-tri\-zed by the affine parameter $t$ of $\ze$ (see Fig.\ $1$).
The region $D$ is thence topologically the
product of  $\Si$ times a real line interval $[t_0,t_1]$. Let the generic leaf be $\Si_t$ and
$B_t$ its boundary, which is obviously $(n-2)$--di\-men\-sio\-nal. Let us then denote by $\vec u$ the
future directed timelike unit normal to the leaf
$\Si_t$. The boundary $\del D$ is formed by  the union of all $\>(n-2)$-- boundaries $B_t$,
which will be denoted by $\calB$, together with the initial and final leaves $\Si_{t_0}$ and
$\Si_{t_1}$ of the sandwich, which will be called {\it lids}.
Let us denote by $\vec n$ the outward spacelike unit normal to $\calB$, by $\ga_{ij}$ the
metric

\leftskip 0truecm
\noindent induced by $g$ on $\calB$, by $h_{ab}$ the metric induced on $\Si_t$ and by $\si_{AB}$
the metric induced on the $(n-2)$--boundary $B_t$.
Here and in the sequel indices run in the following ranges: Greek
indices from $0$ to $n-1$, lower case Roman letters $a,b,\dots$ from $1$ to $n-1$, lower case
middle Roman letters $i,j,\dots$ take the values $0, 2,\dots n-1$ and upper case Roman letters
$A,B\dots$ range from $2$ to $n-1$ (see also Appendix A for the notation).

We shall finally denote by $\Te_{ij}$ the extrinsic curvature of $\calB$ in the spacetime $M$, by
$K_{ab}$ the extrinsic curvature of $\Si_t$ in $M$ and  by $\calK_{AB}$ the extrinsic curvature
of $B_t$ in $\Si_t$ (see Appendix A).
We shall denote by $\Te=\ga^{ij}\>\Te_{ij}$, $K= h^{ab}\> K_{ab}$ and $\calK=\si^{AB}\>\calK_{AB}$
the traces of the  extrinsic curvatures of $\calB$, $\Si_t$ and $B_t$, respectively.
It is assumed that the dynamical metric $g$ and the reference background $\bar g$
induce the same metric $\ga_{ij}$  on $\calB$.
Furthermore, let us also assume that the hypersurfaces $\calB$ and $\Si_t$ intersect
orthogonally (or equivalently that $\vec n$ and $\vec u$ are orthogonal on any $B_t$,
i.e.\ $u^\mu \> n_\mu\vert_{\calB}=0$).
We stress that here we are considering a region $D$ in which the timelike vector field $\ze$
has no fixed points, so that the hypersurfaces $\Si_t$ do not
intersect each other and span the whole region $D$.

According to this notation,  York's action in presence of a background $\bar g$ may be written as
$$
I_D[g,\bar g]= I_D[g]-I_D[\bar g]
\fl{\YAF}$$
where the  functional $I_D[g]$ is defined by:
$$
\eqalign{
I_D[g]=& {1\over 2\ka}\int_{D}\sqrt{g}\> R\> \ds
+{1\over \ka} \int^{\Si_{t_1}}_{\Si_{t_0}}  K\> u^\al\>\sqrt{g}\>\ds_\al
-{1\over \ka} \int_{\calB}   \Te\>n^\al\>\sqrt{g}\>\ds_\al
\cr}
\fl{\YAFO}$$
and $I_D[\bar g]$ is the same functional calculated for the background $\bar g$.
Notice that $\sqrt{g}\>u^\al\>\ds_\al=\sqrt{h}\>\d^3 x$ and
$\sqrt{g}\>n^\al\>\ds_\al=\sqrt{\ga}\>\d^3 x$ are the volume elements on $\Si_t$ and $\calB$,
respectively.
In the functional $\YAFO$ we also set the convenient notation
$\int^{\Si_{t_1}}_{\Si_{t_0}} \equiv \int_{\Si_{t_1}}-\int_{\Si_{t_0}}$.

In the current literature, there exists a whole family of action functionals
similar to $\YAF$ each adapted to the particular problem under consideration.
In general, one can add to the
functional $\YAFO$ an arbitrary functional depending on the data fixed on the boundary, i.e.
depending on the boundary metric (see \ref{\BY}). This arbitrariness  does not affect the
equations of motion since  the boundary metric is kept fixed in the variational principle. The
choice $\YAF$ is motivated by the requirement $I_D[\bar g,\bar g]=0$, i.e. the requirement that the
action functional vanishes when computed with  $g=\bar g$. The same property is satisfied by $\FOL$.

Moreover, under additional hypotheses, the integral on the  lids $\Si_{t_0}$ and $\Si_{t_1}$
in $\YAFO$ is  usually  discarded since  the lids are either ignored  or identified by periodic
boundary conditions (see \ref{\GH}, \ref{\BYdue}).

For these reasons we are forced to first review the literature and choose homogeneous notation to
compare with the covariant first order Lagrangian.

We are not aware of a theoretical comparison between these two variational principles in the current
literature. A definition of energy based on the action
functional $\YAF$ has been proposed by Brown and York. It is called {\it
quasilocal energy} and it is often quoted because it reproduces the ADM mass in the asymptotically
flat case, though it has been recognised to be also suitable for more general boundary conditions
(e.g.\ in asymptotically anti--de--Sitter (see \ref{\BTZRefB}) or asymptotically locally flat
cases; see \ref{\AltriTaubMann}).
The main advantage of Brown--York method is that it allows to define the
energy within a finite region as well as the total one. We remark that the covariant
first order Lagrangian $\FOL$ was originally proposed for exactly the same reasons 
(see \ref{\Katz}, \ref{\Cavalese}).
We also remark that in both cases it has been recognized that {\it absolute conserved
quantities} have a meaning just in particular cases, while for general boundary conditions just
{\it conserved quantities relative to a reference background} are meaningful.
Furthermore, even {\it ``absolute'' conserved quantities} should be interpreted as {\it conserved
quantities relative to some canonical background} (e.g.\ flat Minkowski space in asymptotically
flat spaces).
We also remark that reference backgrounds are particularly relevant in General Relativity,
as well as in other non--linear field theories, since whenever   fields are endowed with a vector
space structure then a canonical choice for the reference exists, namely the {\it zero} section.
If the configuration bundle (see Appendix B for a short  geometric insight into variational
calculus and field theory) is not a vector bundle, as it happens in General Relativity, as well as
e.g.\ in Yang--Mills theories, there is no canonical choice for the vacuum state.
The vacuum state has thence to be arbitrarily fixed.
When doing that it sounds physically reasonable to require also the background to be a solution of
field equations, so that the relative
mass can be interpreted as the energy {\it ``spent to go''} from the background solution to the
dynamical one (and analogously for other currents). Furthermore, it is essential that the choice
of the reference background does not effect the evolution of the dynamical fields, i.e.\ they have
to be decoupled.
 In view of these considerations both action functionals $\FOL$ and $\YAF$ incorporate the
background from the very beginning.
Thus we believe that   the relationship between the two methods is worth
investigating.

In this paper we shall also discuss the very  notion of {\it
conserved quantity}. On one hand, in fact, N\"other theorem provides currents $\calE$ which are 
{\it covariantly conserved}, i.e.\ $\d\calE=0$ identically or on shell (i.e., along solutions),
meaning that their integral on the boundary 
$\del D$ of any $n$--region $D$ in spacetime $M$ vanishes or, equivalently, that the conserved
quantity obeys a continuity equation.
On the other hand, physicists are often interested in quantities which are {\it conserved in time},
meaning that, once an ADM foliation of a region of spacetime has been chosen, such a quantity
$\calQ$ may be computed by integration on each leaf and it turns out not to depend on the
particular leaf labelled by the {\it time $t$}.
Clearly, ADM foliations are far not unique and different foliations of the same region $D$
correspond to different ways of defining time.
Furthermore, such a  quantity $\calQ$ may be conserved in the time
defined by an ADM foliation without being conserved  with respect to other foliations (see
Appendix C). From a theoretical General Relativity viewpoint,  quantities conserved in time are
not (manifestly) covariant in nature.
They are, in fact, conserved with respect to a special parameter, while, at a fundamental level,
the principle of general covariance forbids, at least in principle, the selection of a preferred
time.

Nevertheless,  quantities conserved in time may be interesting to be investigated.
In our perspective, in fact,  they can be obtained from covariantly conserved quantities.
To be more precise, we can consider a variational principle, an infinitesimal generator $\xi$ of
Lagrangian symmetries and a solution of field equations. Then we can compute covariantly
conserved currents $\calE[\xi]$ by N\"other theorem.
Let us then fix a spacelike $(n-1)$--region $\Si$ and integrate the N\"other current on it to
define a  conserved quantity $\calQ[\xi]$.
Any timelike vector field $\ze$ allows then to evolve the region $\Si$ along its flow,
parametrized by its affine parameter $t$.
Under  this viewpoint, the question arises whether there exists a vector field $\ze$ (possibly
depending on the region $\Si$) such that the covariantly conserved quantity generated by $\xi$ is
also conserved in the ``time'' induced by $\ze$.
At a first glance, if we have $\del D=\Si_{t_1}-\Si_{t_0} + \calB $ (see Fig.\ 1), conservation
in time is equivalent to require that the integral of the N\"other current $\calE[\xi]$ on $\calB$
vanishes (for any time interval $[t_0,t_1]$). In fact, we have the covariant conservation law
$\d
\calE[\xi]=0$, so that
$$
\eqalign{
&0=\int_D \d \calE[\xi]=\int_{\del D} \calE[\xi]=\int_{\Si_{t_1}} \calE[\xi]-\int_{\Si_{t_0}}
\calE[\xi]   +\int_{\calB} \calE[\xi]\then\cr
&\then \int_{\Si_{t_1}} \calE[\xi] - \int_{\Si_{t_0}} \calE[\xi]=
-\int_{\calB} \calE[\xi]\cr}
\fn$$  
Thence the conserved quantity $\int_{\Si_{t}} \calE[\xi]$ computed on a leaf does not depend on the
particular leaf if and only if $\int_{\calB} \calE[\xi]=0$ (for any time interval $[t_0,t_1]$).
Physically speaking, this amount to require that the flow  of the current
$\calE[\xi]$ through $\calB$ is vanishing.
Clearly, different ADM foliations may evolve $\Si$ in different ways.
In general, just few of them will lead to time--conserved quantities.
The vanishing of $\int_{\calB} \calE[\xi]$ has then to be guaranteed by additional
hypotheses, possibly in many different ways.
Under stronger hypotheses on $\xi$ (or on $\ze$, or on the boundary conditions which $g$ and
$\bar g$ have to satisfy) the set of ADM foliations leading to time--conserved quantities with
respect to different times may be possibly  enlarged.
Different sets of conditions which  guarantee time--conservation will be discussed below
and we shall compare them to those found in the current literature; see also Appendix C for some
examples.

\ms
In Section $2$ we shall prove that the two action functionals coincide  provided that the
metric $g$ and the background $\bar g$  agree on the boundary $\del D$ of the region $D$ under
consideration.

In Section $3$ the variational principle for  York's action functional will be  reviewed
and the definition of  quasilocal energy  recalled. We shall also briefly review the variational
principle for the first order covariant Lagrangian. The two variational principles are relevant by
their own and of course are found to be equivalent when $g$ and $\bar g$ are required to agree on the
boundary $\del D$, i.e.\ when the action functionals are equivalent.

In Section $4$ a $(3+1)$ decomposition of   York's action functional will be  reviewed.
The obtained   ADM Hamiltonian will be later compared with the N\"other conserved quantity
of the first order covariant Lagrangian as introduced in Section $5$ and $6$.

In Section $5$ we shall review the covariant approach to N\"other conserved quantities and comment
on the state of the art in the definition of the energy (mass, angular momentum, etc.) in General
Relativity. Here most of the geometric techniques and bundle framework have been isolated and
reviewed in Appendix $B$.

In Section $6$ the N\"other theorem will be specialized to define the covariantly conserved
quantities of the first order covariant Lagrangian.
Such quantities are involved in the definition of  the {\it covariant ADM Hamiltonians} (one for each
spacetime vector field $\xi$). They will  in fact be compared with the ADM Hamiltonian introduced
in Section
$4$ and found to agree (under the usual matching conditions required on $g$ and $\bar g$).
This result will be obtained by comparing the two quantities with respect to the same ADM
foliation of the region $D$ under consideration.
Finally, the relation between covariant ADM Hamiltonian and quasilocal energy will  also be 
analysed. The main result is that quasilocal energy is covariantly conserved and it is the
N\"other charge associated to the  unit normal to the leaves of the ADM foliation.

In Section $7$ we shall discuss different sets of conditions under which conservation in time
follows from covariant conservation.

In Appendix A we collect formulae which are used during the paper to translate covariant objects to
objects adapted to the ADM foliation and viceversa.

As we already said, Appendix B contains a quick review of the bundle framework and N\"other theorem
at bundle level.

Finally in Appendix C we present two worked examples which  in various occasions are  quoted
throughout  the paper.
The first one is the computation of various N\"other charges of the Schwarzschild solution relative
to Minkowski metric matched on a finite sphere.
Conservation in  time of various foliations is analysed and quasilocal energy is obtained 
in agreement with previously known results (see \ref{\BY}).
The second example is a Kerr--Newman solution matched at spatial infinity with the  Minkowski
metric. The N\"other conserved quantities are obtained. Such an example is interesting because it
does not obey the same matching conditions required all over the rest of the  paper. 
It nevertheless produces a current which is time--conserved, showing that all the
conditions discussed along the paper are sufficient but not necessary.

\NewSection{Comparison of the Action Functionals}
From now on we assume $\dim (M)=4$, unless explicitly stated.
We shall here decompose the first order covariant action functional $\FOL$  along an ADM foliation 
of $D$ in order to prove  that  the action functional $\YAF$  and $\FOL$ are equal  if the
$4$--metrics  $g$ and $\bar g$ are required to coincide on $\partial D$.

 Of course, the ADM splitting breaks down the
explicit covariance in the action allowing a comparison with respect to the same ADM foliation.
In particular, the boundary term in the covariant action $\FOL$ splits into a contribution on
$\calB$ and a contribution on the {\it lids} $\Si_{t_0}$ and $\Si_{t_1}$.

Let us consider local coordinates $(t,r, x^A)$ adapted both to $\calB$ and the ADM splitting on
$D$. In this coordinate system  $\calB$ has the expression $r=$ {\it constant}  while the leaves
$\Si_t$ are the hypersurfaces of equation $t=${\it constant}.
The metric tensor $g$ can be split with respect to the ADM--foliation obtaining the expression
$(A.13)$ in the Appendix A.
Similarly, one can consider the foliation of spacetime in the hypersurfaces $r=${\it constant}
and obtain the expression $(A.19)$ in the Appendix A.
Analogous expressions can be obtained for the reference background metric $\bar g$.

Let us evaluate the boundary term of the covariant first order action functional $\FOL$ on $\calB$
and on the lids $\Si_{t_0}$ and $\Si_{t_1}$,
i.e.:
$$
A_\calB=-{1\over 2\ka}\int_{\calB}\sqrt{g} g^{\mu\nu} w^\al_{\mu\nu}\> \ds_\al
\qquad\qquad
A_{\Sigma_{t_0}}^{\Sigma_{t_1}}=
-{1\over 2\ka}\int_{\Sigma_{t_0}}^{\Sigma_{t_1}}\sqrt{g} g^{\mu\nu}w^\al_{\mu\nu}\>\ds_\al
\fl{\CovDivergence}$$
By using results which are summarized in  Appendix A (see equations $(A.18)$ and $(A.23)$) one
obtains:
$$
\eqalign{
A_\calB=&-{1\over 2\ka}\int_{\calB} g^{\mu\nu} w^\al_{\mu\nu}\> n_\al \sqrt{\ga}\>\d^3 x
=-{1\over 2\ka}\int_{\calB}\Big\{
2\Te+\cr
&-\bar \Te_{ij}\left( {\bar V\over V\phantom{\big\vert}}\,\bar\ga^{ij}
+{V\over \bar V\phantom{\big\vert}}\,\ga^{ij}\right)
-{\bar\Te_{ij}\over V\bar V\phantom{\big\vert}}\>(V^i-\bar V^i)\>(V^j-\bar V^j)
+\cr
&
+{1\over V\bar V\phantom{\big\vert}}\>\del_i\bar V\>(V^i-\bar V^i)
-{1\over V\phantom{\big\vert}}\>\bar {\cal D}_i(V^i-\bar V^i)
\Big\}\> \sqrt{\ga}\>\d^3 x\cr
}
\fl{\SplittingAzioneB}$$
where $\bar {\cal D}_i$ is the covariant derivative with respect to the $3$--metric $\bar
\ga_{ij}$ induced on $\calB$ by the background metric $\bar g$, while $V$ and $V^i$ are the 
{\it radial lapse}  and the {\it radial shift} as defined  by equation $(A.19)$ in the Appendix A.
We stress that no matching condition is required to obtain the above  result. The behaviours of
the metrics $g$ and
$\bar g$ are completely unrelated till now.

Analogously, on the lids the ADM splitting
of the boundary term $\CovDivergence$ gives an extra
contribution of the following form:
$$
\eqalign{
A_{\Sigma_{t_0}}^{\Sigma_{t_1}}=&+{1\over 2\ka}\int_{\Sigma_{t_0}}^{\Sigma_{t_1}}
g^{\mu\nu} w^\al_{\mu\nu}\> u_\al
\sqrt{h}\>\d^3 x 
= - 
{1\over 2\ka}\int_{\Sigma_{t_0}}^{\Sigma_{t_1}}\!\Big\{
\!-2 K+ \cr
&+\bar K_{ab}\left(
{\bar N\over N}\,\bar h^{ab}
+{N\over\bar N}\,h^{ab}\right) 
-{\bar K_{ab}\over N\bar N}\>(N^a-\bar N^a)\>(N^b-\bar N^b)+\cr 
&
-{1\over N\bar N}\>\del_a \bar N\>(N^a-\bar N^a)
+{1\over N}\>\bar D_a(N^a-\bar N^a)
\Big\}\> \sqrt{h}\>\d^3 x\cr
}
\fl{\SplittingAzioneLids}$$
where $\bar D_a$ is the covariant derivative with respect to the $3$--metric $\bar
h_{ab}$ induced on $\Sigma$ by the background metric $\bar g$, while $N$ and $N^a$ are the 
 lapse  and the   shift of the metric as defined  by equation $(A.13)$ in the Appendix A.
Once again we stress that no matching condition is required to obtain the result.
In general (i.e. if no  physical requirement about the matching of
the dynamical metric and the background on the boundary is imposed) the two action functionals 
$\FOL$ and  $\YAF$ are fairly different.

However, let us assume that the dynamical metric $g$ and the background $\bar g$ 
coincide on the hypersurface $\calB$ so that, in particular, they 
induce the same
$3$--metric on $\calB$ (i.e.\
$\ga_{ij}\left.\right\vert_{\calB}=\bar\ga_{ij}\left.\right\vert_{\calB}$) and  they have the
same {\it radial} lapse function (i.e.\ $V\left.\right\vert_{\calB}=\bar V\left.\right\vert_{\calB}$)
and {\it radial} shift vector (i.e.\ $V^i\left.\right\vert_{\calB}=\bar
V^i\left.\right\vert_{\calB}$). Then, under these additional  hypotheses, the contribution
$A_\calB$ reduces to:
$$
A_\calB=-{1\over \ka}\int_{\calB} \Big(\sqrt{\ga}\>\Te -\sqrt{\bar\ga}\>\bar \Te\Big) \>\d^3 x
=-{1\over \ka}\int_{\calB} \Big(\sqrt{g} \Te\> -\sqrt{\bar g}\bar \Te\Big)\>n^\al\>\ds_\al
\fn$$
Analogously, if the metric $g$ and $\bar g$ are required to agree on the lids
(i.e.\ if $h_{ij}=\bar h_{ij}$, $N=\bar N$ and $N^i=\bar N^i$ on $\Sigma_{t_0}$ and $\Sigma_{t_1}$),
then the contribution $\SplittingAzioneLids$ on the lids reduces to:
$$
A_{\Sigma_{t_0}}^{\Sigma_{t_1}}={1\over \ka}\int_{\Sigma_{t_0}}^{\Sigma_{t_1}} \Big(\sqrt{h}\>K
-\sqrt{\bar h}\>\bar K\Big) \>\d^3 x= {1\over \ka}\int_{\Sigma_{t_0}}^{\Sigma_{t_1}}
 \Big(\sqrt{g}\> K\> -\sqrt{\bar g}\>\bar K\Big)\>u^\al\>\ds_\al
\fn$$
Then the boundary term in the covariant first order action $\FOL$
can be written as:
$$
A_\calB + A_{\Sigma_{t_0}}^{\Sigma_{t_1}}=
-{1\over \ka}\int_{\calB} \Big(\sqrt{g} \Te\> -\sqrt{\bar g}\bar \Te\Big)\>n^\al\>\ds_\al
+{1\over \ka}\int_{\Sigma_{t_0}}^{\Sigma_{t_1}}
 \Big(\sqrt{g}\> K\> -\sqrt{\bar g}\>\bar K\Big)\>u^\al\>\ds_\al
\fn$$
and the action functionals $\FOL$ and $\YAF$ clearly concide.

We stress that this result has been obtained by requiring the aforementioned  matching conditions
between
$g$ and
$\bar g$ on the complete boundary $\partial D=\calB+\Si_{t_1}-\Si_{t_0}$ of the region $D$.

We should also remark that for time--independent  solutions the matching on the lids cannot be
required, because if the two  metrics $g$ and $\bar g$ agree on a spacelike hypersurface $\Si_t$,
they necessarily agree on the whole region $D$. Also for this reason,  the
contributions on the lids are never considered in applications;
this is usually done by restricting to situations in which the lids are not present
(e.g.\ by considering a non--compact region $D$ in which $t_0$ and $t_1$ are let to tend to
$-\infty$ and $\infty$, respectively) or are identified.
[Identification is obtained, e.g., when the solution is time--periodic, as it may happen in the
Euclidean sector (see \ref{\GH}, \ref{\BYdue}, \ref{\TaubNUTH}) and  in approaches based
on path--integrals for evaluating the  grand--canonical partition function or the density
of states for General Relativity, where the sum over periodic histories has to be considered
(see \ref{\BYdue}).
In all those cases the boundary $\partial D$ is {\it required} to have the topology $\del\Si\times
S^1$, i.e.\ it is assumed to be a single boundary component $\del D=\calB$.]

The matching condition of the $4$--metrics $g$ and $\bar g$ is a  stronger requirement than the one
introduced   in \ref{\HawHun}. There, only the induced  $3$--metrics $\ga_{ij}$ and  $\bar
\ga_{ij}$ are required to agree on $\calB$, where $\calB$ is  let to tend to {\it infinity}. 
Although the matching  of the $4$--metrics at infinity is not too hard to be implemented 
in applications (see \ref{\Cavalese}, \ref{\BTZ}, \ref{\TaubNUTH}),  we are aware that  this
matching, when possible,
 may become hard to implement in a {\it finite region}.
Here the matching is required  to have a direct theoretical comparison between the action functionals
$\FOL$ and $\YAF$ (and between their Hamiltonians and conserved charges).
In Appendix C we shall explicitly discuss a simple but   relevant example of the  matching of the
$4$--metrics in a finite region.

\NewSection{Comparison of the Variational Principles}

We shall here review the variational principles associated to the action functionals $\FOL$ and
$\YAF$. They are found to agree when the  matching conditions already discussed in Section $2$ are
again imposed. We shall also recall the definition of quasilocal energy which will be later
compared with N\"other conserved quantities.

Let us analyse  the variation of the action functional $\YAF$. It reads as (see \ref{\York}):
$$
 \de I_D[g]=\fraz 1{2\ka}\int_D G_{\mu\nu}\,\delta g^{\mu\nu}\,
ds+\int_{\calB}  \,(\Pi^{ij}\,\delta \gamma_{ij})\,d^3x
         +\int_{\Sigma_{t_0}}^{\Sigma_{t_1}}  \,(P^{ab}\,\delta h_{ab})\,d^3x
\fl{\deltagby}
$$
where $G_{\mu\nu}=R_{\mu\nu}-\fraz 12 \,R\, g_{\mu\nu}$ is the Einstein tensor,
while
$\Pi^{ij}$ and $ P^{ab }$ denote the {\it gravitational momenta} conjugated to the metrics
$\gamma_{ij}$ and $h_{ab}$, respectively; they are given by:
$$
\left\{
\eqalign{ 
&\Pi^{ij}=-\fraz 1 {2\ka}\sqrt{\gamma}\,(\Te\,\ga^{ij}-\Te^{ij})\cr
& P^{ab}= \fraz 1 {2\ka}\sqrt{h}\,(K\,h^{ab}-K^{ab})\cr
}\right. 
\fl{\momentiP}
$$
Analogous terms come from the variation $\de I_D[\bar g]$.
Notice that, if boundary conditions which keep the $3$--metric fixed on the boundary are
imposed, namely $\delta\ga\vert_{\calB}=0$ and
$\delta h\vert_{\Sigma_{t_0}}=\delta h\vert_{\Sigma_{t_1}} =0 $,
then the action functional $\YAF$ is correctly extremized by the solutions of Einstein equations
$G_{\mu\nu}=0$.
[We recall that to obtain the  expression ${\deltagby}$ the hypothesis of
orthogonal boundaries has been assumed. Otherwise further terms would appear which have to be
cancelled through the introduction of a further boundary term into the action $\YAF$;
see \ref{\Haywardnon}, \ref{\Mannultimo}, \ref{\HawHun}.]
\medskip
The contribution $\delta I_\calB=\int_{\calB} \,(\Pi^{ij}\,\delta \gamma_{ij})\,d^3x $ in
$\deltagby$  is considered (see \ref{\BY}) in order to define the quasilocal energy $E[g]$, i.e.
the energy of a region of finite spatial extent:
$$
E_t[g]= -\int_{B_t} {\delta I_\calB\over \delta N}\,d^2 x 
\fl{\qlef}
$$
It represents the change of the action in time, where changes in time are governed by
the lapse $N$ on the boundary $\calB$. We stress that the  interpretation of this
quantity as the  energy of a gravitating system was suggested in  \ref{\BY}  through a
Hamilton--Jacobi analysis of the action functional.

Inserting into $\deltagby$  the following decomposition (see \ref{\BY} for the details):
$$
\delta \ga_{ij}=-2 {u_i u_j\over N}\,\delta N -2 {\si_{k(i} u_{j)}\over N}\, \delta N^k+
\si^k_{i} \si^h_{j}\, \delta \si_{kh}
\fn
$$
we obtain:
$$
E_t[g]=2\int_{B_t} \left({\Pi^{ij}\over N}\, u_i u_j\right)\, d^2 x={1\over \ka}\int_{B_t}
 \calK\>\sqrt{\si}\,d^2 x
\fl{\qlocalener}
$$
(in the latter equality we have used the definition $\momentiP$ together with the
expression $(A.11)$ and $(A.12)$ in the Appendix A). A contribution analogous to $\qlocalener$
comes from the background metric
$\bar g$ and from the variation 
$\delta I_\calB[\bar g]$. Assuming that the induced  $3$--metric $\ga$ and $\bar \ga$ are matched
on the surface $\cal B$ we finally have:
$$
E_t[g,\bar g]={1\over \ka}\int_{B_t}
(\calK -\bar{\calK})\>\sqrt{\si}\,d^2 x
\fl{\qletotale}
$$
This is  the explicit expression of Brown and York quasilocal energy computed for the action
functional $\YAF$.
\medskip

Let us now consider the action functional $\FOL$.
We can perform the {\it covariant variation} with
respect to the metrics before choosing any ADM splitting.
In this way we obtain:
$$
\eqalign{ 
\de A_D&[g,\bar g]=\fraz 1{2 \ka}\int_D G_{\mu\nu}\> \de g^{\mu\nu}\,\ds
-\fraz 1{2 \ka}\int_D \bar G_{\mu\nu}\>\de \bar g^{\mu\nu}\,\ds+\cr
&+\fraz 1{2\ka}\int_{\del D}\Big(\sqrt{g}\> g^{\mu\nu}\>\de u^\al_{\mu\nu}
-\sqrt{\bar g}\> \bar g^{\mu\nu}\>\de \bar u^\al_{\mu\nu}
-\de(\sqrt{g}\> g^{\mu\nu}\> w^\al_{\mu\nu})\Big)\>\ds_\al=\cr
=&\fraz 1{2\ka}\int_D G_{\mu\nu}\, \delta g^{\mu\nu}\,\ds
-\fraz 1{2 \ka}\int_D \bar G_{\mu\nu}\,\de \bar g^{\mu\nu}\,\ds+\cr
&+\fraz 1{2 \ka}\int_{\del D}\Big(\big(
\sqrt{g}\, g^{\mu\nu}-\sqrt{\bar g}\, \bar g^{\mu\nu}
\big)\>\de \bar u^\al_{\mu\nu}
-\delta(\sqrt{g}\, g^{\mu\nu})\, w^\al_{\mu\nu}\Big)\,\ds_\al \cr
}  
\fl{\variazcov}$$
Then, if both $g$ and $\bar g$ are solutions of Einstein equations $G_{\mu\nu}=0$ and
$\bar G_{\mu\nu}=0$, they extremize the action $\FOL$, provided that
$\de g^{\mu\nu}
=0$ and $\de \bar u^\al_{\mu\nu}=0$ on the boundary
$\del D$.

This is a stronger requirement than the fixing of the metric induced on $\del D$.
In fact, it amounts to fix the whole $4$--metric restricted on $\del D$ (as well as the first
order derivatives of the reference metric $\bar g$). In any case it is what can be done in order
to preserve covariance if no additional hypotheses on the background are required.

Nevertheless, we again stress that the action $\FOL$ is extremized by solutions of
Einstein equations  also if we require $\left.\delta g^{\mu\nu}\right\vert_{\partial
D}=0$ together with the matching condition $\left.g^{\mu\nu}\right\vert_{\partial
D}=\left.\bar g^{\mu\nu}\right\vert_{\partial D}$. 
This is obvious from expression
$\variazcov$ but it also follows from the demonstration carried out  in the previous
Section: under the assumption
$\left.g_{\mu\nu}\right\vert_{\partial
D}=\left.\bar g_{\mu\nu}\right\vert_{\partial D}$  the action functional $\FOL$ equals
the action functional $\YAF$. Hence the variation $\variazcov$ turns out to coincide with
the variation $\deltagby$.

\NewSection{ADM Hamiltonian}
We shall here review the $(3+1)$ Hamiltonian formulation of the action functional $\YAF$. The ADM
Hamiltonian will be later compared with N\"other charges.

 As it is well known, one of the possible way to give a Hamiltonian description of
General Relativity is the so--called ADM formulation (see \ref{\ADM}, \ref{\Gravitation},
\ref{\II}). We remark that in this framework the concept of (manifest) general covariance is lost
from the beginning owing to the ADM splitting of spacetime into space$+$time. We shall see below
how a covariant ADM formulation  of the problem, deeply related to conserved quantities and
N\"other theorem, can be formulated.  But, first of all, in order to compare known results with
ours on the subject
 let us review the standard ADM  Hamiltonian formulation of the action $\YAF$.

To obtain the   Hamiltonian of the system we
need to separate out the terms in the volume integral  $\fraz 1{2\ka}\int_D
\sqrt{g} R\,\ds$  which are pure divergences and then become boundary
integrals.
Using the decomposition $(A.13)$ of the metric we have (see  \ref{\Gravitation}):
$$
\sqrt{g} R=\sqrt{g}\left({\cal R}+K_{\mu\nu}K^{\mu\nu}-K^2   \right)-2 \sqrt{g}\,
\nabla_\mu\left( K  u^\mu+a^\mu\right)
\fn
$$
where  ${\cal R}$ is the scalar curvature of the $3$--metric $h_{\mu\nu}$ while
$a^\mu=u^\nu\nabla_\nu u^\mu$ is the covariant acceleration of the unit normal $ u^\mu$.
Inserting this  expression into $\YAFO$ we obtain:
$$
\eqalign{
I_D[g]=& {1\over 2\ka}\int_{D}\sqrt{g}\left({\cal R}+K_{\mu\nu}K^{\mu\nu}-K^2  
\right)\ds\cr 
 & +{1\over \ka} \int^{\Si_{t_1}}_{\Si_{t_0}}u_\mu(K u^\mu+a^\mu)\sqrt{h}\, d^3x
-{1\over \ka} \int_{\calB}n_\mu(K u^\mu+a^\mu)\sqrt{\ga} d^3x\cr
&  +{1\over \ka} \int^{\Si_{t_1}}_{\Si_{t_0}}  K  \sqrt{h} \,d^3x
-{1\over \ka} \int_{\calB} \Theta  \sqrt{\ga} \,d^3x
\cr}
\fl{\aziointer}
$$
Owing to the properties  $u_\mu\,u^\mu=-1$ and $u_\mu\,a^\mu=0$, the integral   
$\int^{\Si_{t_1}}_{\Si_{t_0}}$ on the lids vanishes. Moreover, since we have imposed the
condition 
$\left.n_\mu\,u^\mu\right\vert_{\calB}=0$ of orthogonal boundaries, the action functional $I_D[g]$
reduces to:
$$
I_D[g]= {1\over 2\ka}\int_{D}\left({\cal R}+K_{\mu\nu}K^{\mu\nu}-K^2  
\right)\>\sqrt{g}\,\ds
-{1\over \ka} \int_{\calB}\left(\Theta+n_\mu\,a^\mu\right)\sqrt{\ga}\, d^3x
\fn
$$ 
By using formula $(A.12)$ in the Appendix A we obtain:
$$
I_D[g]= {1\over 2\ka}\int_{D}\left({\cal R}+K_{\mu\nu}K^{\mu\nu}-K^2  
\right)\>\sqrt{g}\,\ds
-{1\over \ka} \int_{\calB}\,{\cal K}\sqrt{\ga}\, d^3x
\fn
$$
This  action may be written in canonical form by inserting the
equality:
$$
K_{ab}K^{ab}-K^2= {2\ka\over{\sqrt{g}}}\left\{P^{ab}\partial_t h_{ab}-2 P^{ab}D_a N_b-
{\ka N\over{\sqrt{h}}}\left(2P^{ab}P_{ab}-P^2
\right)\right\}
\fn
$$
(see definition $\momentiP$) and by removing the term involving the derivatives $D_a N_b$ by 
an integration by parts. We finally have:
$$
\eqalign{
I_D[g]=&\int\! dt\int_{\Sigma_t}\left(P^{ab} \partial_t h_{ab}-N{\cal H}-N_a
{\cal H}^a\right)\!d^3x+\cr
-&\int \!dt\int_{B_t}\left({N\over {\ka}}\,{\cal K}+N^a\,\,2{P^{bc}\over
{\sqrt{h}}}\, n_b\si_{ac}
\right)\>\sqrt{\si}\!d^2x\cr
}
\fl{\azionetrepiuuno}
$$
where
$$
\eqalign{
&{\cal H}=-{1\over 2\ka}\sqrt{h}\,{\cal R} +{\ka\over \sqrt{h}}(2 P_{ab} P^{ab}-P^2)    \cr
&{\cal H}^a=-2 D_b P^{ab}\cr
}
\fn
$$
are the Hamiltonian   and the momentum constraint, respectively. The  Hamiltonian
for vacuum General Relativity is  thus identified with the term:
$$
H(g)=\int_{\Sigma_t}\left(N{\cal H}+N_a
{\cal H}^a\right)\!d^3x+
{1\over {\ka}}\int_{B_t}\left(N\,{\cal K}-N^a\,\,K^{bc}\,
n_b\si_{ac}
\right)\>\sqrt{\si}\!d^2x
\fl{\ADMHamilt}$$
(in the latter term we have taken into account that $h^{bc}n_b\si_{ac}=0$).
The Hamiltonian $H(g)$ is the sum of a {\it constrained volume} term, which is vanishing when computed
on a solution, and a {\it boundary} term.

We may repeat the above analysis  also for the background action functional $I_D[\bar g]$
obtaining an Hamiltonian $H(\bar g)$ which agrees with  $\ADMHamilt$ provided that we replace the
terms there involved  with the corresponding barred ones. 
The total Hamiltonian $H(g,\bar g)$ is then given by the difference:
$$
H(g,\bar g)=H(g)-H(\bar g)
\fl{\HmenoHbar}
$$
When it is evaluated on solutions of field equations,  the constrained volume terms 
 vanish and $\HmenoHbar$ reduces  to the boundary terms:
$$
\eqalign{
H(g,\bar g)&\simeq {1\over {\ka}}\int_{B_t}\left(N\,{\cal K}
-N^a\,\,K^{bc}\, n_b\si_{ac}\right)\>\sqrt{\si}\!d^2x\cr
&-{1\over {\ka}}\int_{B_t}\left(\bar N\,\bar{\cal K}
-\bar N^a\,\,\bar K^{bc}\, \bar n_b\bar\si_{ac}\right)\>\sqrt{\bar\si}\!d^2x\cr
}
\fl{\Hgbarg}
$$
where the symbol $\simeq$ denotes  equality on--shell.  

Moreover, if the metric $g$ and its relative background $\bar g$ agree on the boundary the
Hamiltonian simplifies as follows:
$$
H(g,\bar g)\simeq {1\over {\ka}}\int_{B_t}\left\{N\,({\cal
K}-\bar {\cal K})-N^a\,(K^{bc}-\bar K^{bc})\, n_b\si_{ac}\right\}\>\sqrt{\si}\!d^2x
\fl{\ADMHamilton}
$$
The {\it mass} associated with the time
translation
$t^\mu=N u^\mu+ N^\mu$, relative to  two solutions $ g$ and $\bar g$, is simply defined  to be 
the value of the Hamiltonian $\Hgbarg$.
Clearly the mass $H(\bar g,\bar g)$ of the background is equal
to zero. 

Notice that, choosing a Gaussian gauge, i.e. setting $N=1$ and $N^a=0$ in $\ADMHamilton$, we
obtain the quasilocal energy $\qletotale$ (see \ref{\BY}). 

In \ref{\HawHor} it was shown that the definition  $\Hgbarg$ agrees with the  expressions  of
energy already defined  in literature for spacetimes with different asymptotic behaviour.

\NewSection{Conserved Quantities}

In this Section we shall analyse conserved quantities associated to the covariant first order
Lagrangian $\FOL$ by the N\"other theorem.
As we claimed above, the Lagrangian $\FOL$ was originally introduced because it provides a simple
framework to determine the density of conserved quantities.
We shall here apply the general framework (see e.g.\  \ref{\Katz}, \ref{\Cavalese},
\ref{\BTZ}, \ref{\Remarks} and references quoted therein).

The N\"other theorem is a direct consequence of the general covariance of General Relativity  (as
well as of any other generally covariant field theory).
It {\sl algorithmically} defines a {\it N\"other conserved current}, i.e.\ a map $\calE[\xi,\si]$
which associates to any spacetime vector field $\xi$ and any field configuration $\si=(g,\bar g)$
an  $(n-1)$--form on  spacetime $M$ of dimension $n$ which is closed {\sl on--shell}, i.e.\ when the
configuration
$\si$ is a solution of field equations.
For the action functional $\FOL$ we obtain:
$$
\eqalign{
\calE[\xi,\si]=&{1\over 2\ka}\Big[
\sqrt{g}\>\Big((g^{\la\al}g_{\mu\nu}-\de^{(\la}_\mu\de^{\al)}_\nu)\> \na_\al\Lie_\xi g^{\mu\nu}
-\xi^\la\>R\Big)+\cr
&-
\Big(\Lie_\xi(\sqrt{g}\>g^{\mu\nu}\> w^\la_{\mu\nu})
-\xi^\la\>\d_\al(\sqrt{g}\>g^{\mu\nu}\> w^\al_{\mu\nu})\Big)+\cr
&-
\sqrt{\bar g}\>\Big((\bar g^{\la\al}\bar g_{\mu\nu}-\de^{(\la}_\mu\de^{\al)}_\nu)\> \bar
\na_\al\Lie_\xi \bar g^{\mu\nu} -\xi^\la\>\bar R\Big)
\Big]\>\ds_\la\cr
}
\fl{\NotherCurrent}$$
The differential of the $(n-1)$--form $\calE[\xi,\si]$ satisfies the following property:
$$
\d\calE[\xi,\si] = \calW[\xi,\si]
\fl{\WeakConservation}$$
where
$\calW[\xi,\si]=-(1/2\ka)\big(G_{\mu\nu}\> \Lie_\xi g^{\mu\nu}-  \bar G_{\mu\nu}\> \Lie_\xi\bar
g^{\mu\nu}\big)\>\ds$ is proportional to field equations.
Consequently, the N\"other current $\calE[\xi,\si]$ is closed along solutions.

The equation $\WeakConservation$ is called {\it weak conservation law} for the N\"other current
$\NotherCurrent$.
We stress that equation $\WeakConservation$ is written on  spacetime $M$, where 
it  does not single out a unique N\"other current $\calE[\xi,\si]$.
In fact, if a closed form is added to $\calE[\xi,\si]$ we get another solution of
$\WeakConservation$.
This is why N\"other's theorem should be regarded as a claim on (a suitable prolongation of) the
configuration bundle (see Appendix B), i.e.\
$\calE[\xi,\si]$ has to be regarded as a map $\calE[\xi]$ which associates
a form on (the jet prolongation of) the configuration bundle to any spacetime vector field $\xi$.
As shown in \ref{\Remarks} and \ref{\Robutti}, the N\"other current $\calE[\xi]$ at bundle level is
{\it canonically} associated to the Lagrangian.
{\sl Then} it can  be computed along a configuration $\si$ to  give $\calE[\xi,\si]$, i.e. a
$(n-1)$--form on the spacetime $M$.

Of course, the N\"other currents, as well as the conserved quantities associated to them,
explicitly depend on boundary terms in the Lagrangian.
This is a very well known feature in Physics, as it can be simply seen, e.g., in thermodynamics. It
is in fact well known that boundary terms are related to boundary conditions. As we remarked in 
Section $4$, different boundary terms need different boundary conditions to keep field equations
satisfied by action extremals (see \ref{\York}, \ref{\BYdue}, \ref{\Nester}). And it is well known
that, e.g., for a correct definition of energy in thermodynamics, different boundary conditions
correspond to different definitions of energy, such as {\it internal energy}, {\it free energy} etc.
We stress that all these energies are {\it true} physical energies of thermodinamical systems.
Which one is to be used in practice is determined by the particular system under consideration
and the boundary conditions {\it we decided} to impose.
As it is physically relevant  to notice,  we may decide to keep temperature fixed on the boundary of
a gas box or we may impose adiabatic conditions; this different choice corresponds to a different
apparatus which selects a different energy flow through the boundary so that the boundary conditions
are satisfied. We stress that this corresponds to an external action on the system which turns out to
change the physical energy of the system itself.
In the covariant first order approach to General Relativity something fully analogous holds: the
background
$\bar g$ canonically selects both the boundary conditions and the corresponding energy to be used
and different choices of the background correspond to different physically meaningful definition
of energy.

One can also define (see Appendix B) the {\it superpotential $\calU[\xi,\si]$} and the {\it reduced
current
$\tilde\calE[\xi,\si]$} as those currents such that:
$$
\calE[\xi,\si]=\tilde\calE[\xi,\si]+\d\>\calU[\xi,\si]
\fl{\NotherDecomposition}$$
where the reduced current is required to vanish on--shell.

Once again both $\tilde\calE[\xi,\si]$ and $\calU[\xi,\si]$ are not uniquely identified
by equation $\NotherDecomposition$.
The superpotential is defined modulo forms which are closed on--shell, while the  reduced current
is defined modulo forms vanishing along solutions.
The decomposition $\NotherDecomposition$ of N\"other's current is again well--defined only at
bundle level, where a decomposition algorithm can be constructed (see \ref{\Robutti},
\ref{\Remarks}). The bundle superpotential $\calU[\xi]$ and the reduced current $\tilde\calE[\xi]$
are canonically and globally defined and {\sl then} they are computed along the configuration
$\si$ to give
$\calU[\xi,\si]$ and $\tilde\calE[\xi,\si]$, respectively.

The {\it conserved quantity in a region $\Omega$} is defined as:
$$
Q_\Omega[\xi,\si]=\int_\Omega \calE[\xi,\si]= \int_{\Omega} \tilde\calE[\xi,\si]+
\int_{\del \Omega} \calU[\xi,\si]=\int_{\del \Omega} \calU[\xi,\si]
\fl{\ConservedQuantity}$$
where, in the last equality, $\si$ is assumed to be a solution so that $\tilde\calE[\xi,\si]=0$.
\medskip

In the case of the first order covariant action functional $\FOL$ we obtain in particular:
$$
\eqalign{
&\calU[\xi]={1\over 2\ka}\Big[
\sqrt{g}\> \na^\be_{\>\cdot}\xi^\al +\sqrt{g}\>g^{\mu\nu}\>w^\be_{\mu\nu}\xi^\al-
\sqrt{\bar g}\> \bar\na^\be_{\>\cdot}\xi^\al
\Big]\>\ds_{\al\be}\cr
&\tilde\calE[\xi]={1\over \ka}\Big[
\sqrt{g}\>g^{\mu\la} \>G_{\mu\nu}\>\xi^\nu
-\sqrt{\bar g}\>\bar g^{\mu\la} \>\bar G_{\mu\nu}\>\xi^\nu
\Big]\>\ds_{\la}\cr
}
\fl{\Superpotentials}$$
where $g^{\mu\nu}$ (as well as Christoffel's symbols $\Ga^\la_{\si\mu}$) has to be regarded as local
coordinates on (the jet prolongation of) the configuration bundle $\Lor(M)$ of all Lorentzian
metrics on $M$. They become the metric components, i.e.\ functions of spacetime point $x\in M$, only
when they are calculated along a configuration $\si(x)=(g(x), \bar g(x))$.

The conserved quantity $Q_\Omega[\xi,\si]$ depends on the dynamical metric $g$ and on the
reference background metric $\bar g$.
It has to be interpreted as the {\it relative conserved quantity of $g$ with respect to $\bar g$}.
For example, if the energy is considered (by choosing the vector field $\xi$ in a suitable way, see
below),
$Q_\Omega[\xi,\si]$ represents the amount of energy which is necessary to {\it pass} from the
spacetime
$(M,\bar g)$ to the spacetime $(M,g)$. As it is physically reasonable, this energy can be infinite
in principle. Of course, if the metrics are {\it ``near''} the energy between them can be
expected to be finite, though we cannot even be sure that we can {\it ``continuously''} join two
very different metrics. In general we can expect conserved quantities to classify accessibility
classes of metrics, though we do not want here to enter the problem of providing the set of
sections of the configuration bundle with a topology or a differentiable structure.
We certainly expect that if $\si=(\bar g,\bar g)$, i.e.\ if we set $g=\bar g$, the conserved
quantity $Q_\Omega[\xi,\si]$ vanishes.
This condition is satisfied by prescription $\ConservedQuantity$ because of the form of the
superpotential $\Superpotentials$.

We remark that the conserved quantities defined by $\ConservedQuantity$ are {\it covariantly
conserved}, meaning that their flows through boundaries $\partial D$ of $4$--dimensional regions
$D$ in spacetime $M$ identically vanish.
If one chooses $g$ to be a solution asymptotically flat according to one of the current
definitions (e.g.\ the Kerr--Newman solution), $\bar g$ to be the flat reference background
(which matches the dynamical metric at infinity where ``infinity'' is prescribed by the
definition of asymptotic flatness) and
$\Omega$ to be a spacelike hypersurface in
$M$, then
$Q_\Omega[\xi,\si]$ reproduces the expected value for mass (by chosing $\xi=\del_t$, i.e. the
vector field which corresponds to asymptotic time translation) and angular momentum (by choosing
$\xi=-\del_\phi$, i.e.\ the vector field which corresponds to
asymptotic rotation; see e.g.\ \ref{\KatzBondi}, \ref{\Cavalese}, \ref{\Remarks}).
Furthermore, the same results are achieved for non--asymptotically flat solutions by
choosing suitable reference backgrounds.
Examples are the $(2+1)$ BTZ solution (which is asymptotically anti--de--Sitter, see \ref{\BTZ}),
the Euclidean Taub--Bolt solution (which is asymptotically locally flat, see \ref{\TaubBolt}).
In addition, the same techniques are used successfully in {\it gauge--natural theories}, i.e.\
when the field theory owns both covariance and  gauge invariance (e.g.\ BCEA theory,
Einstein--Maxwell theory;  see \ref{\Remarks}, \ref{\BCEA}). 

We stress that the conserved quantities $Q_\Omega[\xi,\si]$ associated to the covariant first order
action principle $\FOL$ {\it are not} affected by the anomalous factor problems as the ones
associated to the standard Hilbert--Einstein Lagrangian (see \ref{\Katz}).
As is well known, in fact, the Komar superpotential (see \ref{\Komar}):
$$
\calU_{_{K}}={1\over 2\ka}\sqrt{g}\> \na^\be_{\>\cdot}\xi^\al \>\ds_{\al\be}
\fn$$
which is the superpotential associated to the Hilbert--Einstein Lagrangian, when computed, for
example, on the Kerr--Newman solution, produces the correct angular momentun, but just one--half
of the expected mass.
Of course one could postulate the mass to be associated to the vector field $\xi'=2\del_t$.
However, the factor appears to depend on the particular solution under investigation, thus such a
prescription seems to be incorrect as well as unmotivated.
This is usually interpreted as a hint of a correction needed in the definition of conserved
quantities. The correction can be  obtained by the ADM techniques (see \ref{\ADM})
by restricting to asymptotically flat solutions and by choosing an ADM foliation of  spacetime
$M$. As shown in  \ref{\Sinicco}, \ref{\CADM}, \ref{\CADMC} and Section $6$, the ADM formalism
appears as a particular case of the technique exposed above, which furthermore applies to much
more general situations.

\NewSection{Covariant ADM Hamiltonian}
We shall here specialize  the general framework introduced in Section $5$ to the first order action
functional $\YAF$. We thence obtain  a map $Q[\xi]$ which associate to each spacetime vector field a
covariantly conserved quantity, called {\it the covariant } ADM Hamiltonian. It will be compared with
the standard ADM Hamiltonian introduced in Section $4$ and with the quasilocal energy  defined in
Section $3$.

Let us consider the covariant conserved quantity $Q_{\Sigma_t}[\xi,\si]$ for the first
order action functional
$\FOL$ in the domain
$\Sigma_t$ and relative to a vector field $ \xi$ and a section $\si=(g,\bar g)$. As we have
already outlined in the previous Section, the quantity  $Q_{\Sigma_t}[\xi,\si]$ is defined to be
the integral of the superpotential $\Superpotentials$ on the $2$--dimensional surfaces $B_t$
(see Fig.\ $1$ and Appendix A for the notation),
 i.e.
$$
Q^{\hbox{\fiverm Tot}}_{\Sigma_t}[\xi,\si]=Q_{\Sigma_t}[\xi,g]+Q_{\Sigma_t}[\xi,g,\bar
g]+Q_{\Sigma_t}[\xi,\bar g]
\fl{\qsigmat}$$
where:
$$
\eqalign{
&Q_{\Sigma_t}[\xi,g]={1\over 2\ka}\int_{B_t}\sqrt{g}\>
        \na^\be_{\>\cdot}\xi^\al\ds_{\al\be}\cr
&Q_{\Sigma_t}[\xi,g,\bar g]={1\over
2\ka}\int_{B_t}\sqrt{g}\>g^{\mu\nu}\>w^\be_{\mu\nu}\xi^\al\ds_{\al\be}\cr
&Q_{\Sigma_t}[\xi,\bar g]=-{1\over 2\ka}\int_{B_t}\sqrt{\bar g}\>
        \bar \na^\be_{\>\cdot}\xi^\al\ds_{\al\be}\cr
}
\fl{\treQ}
$$
In order to simplify the  ADM decomposition of the expression $\qsigmat$ so  to be  
able to compare the results obtained with the standard ones  of Section $4$ and
\ref{\BY} let us assume, as usual, that the  metrics $g$ and $\bar g$ are matched on
the hypersurface $\calB$ and that the boundaries are orthogonal (i.e.
$\left. u^\mu n_\mu\right\vert_{\calB}=0$). 

[We stress that under  our viewpoint the matching condition between $g$ and $\bar g$ is
unessential, since N\"other currents are covariantly conserved. One may consider the
second example analysed in Appendix C where the Kerr solution  is studied and its {\it mass }
inside the finite R--sphere is obtained with respect to a flat background matched at infinity. 
In this Section we require the matching on
$\calB$ in order  to compare the N\"other charges expression with the aforementioned standard
$(3+1)$ Hamiltonian and quasilocal energy.]

Let us also assume that the vector field
$\xi$ is tangent to the hypersurface
$\calB$ (i.e. 
$\left. \xi^\mu n_\mu\right\vert_{\calB}=0$).
First of all let us consider the first contribution $Q_{\Sigma_t}[\xi,g]$ into
$\qsigmat$, i.e. the integral of the Komar superpotential. It may be rewritten as:
$$
Q_{\Sigma_t}[\xi,g]={1\over 2\ka}\int_{B_t} (u_\be n_\al-u_\al
n_\be)\,g^{\be\mu}\,\nabla_\mu\xi^\al\>\sqrt{\si}\,d^2x
\fn$$
where $\sqrt{\si}\,d^2x,$ is the volume element on $B_t$. We can manipulate algebraically
the latter expression in the following way:
$$
\eqalign{
Q_{\Sigma_t}[\xi,g]&={1\over 2\ka}\int_{B_t}
\left\{g^{\be\mu}\,\nabla_\mu\xi^\al(2u_\be\,n_\al-u_\be\, n_\al-u_\al\,
n_\be)\right\}\,\sqrt{\si} d^2 x=\cr
&= {1\over 2\ka}\int_{B_t}
\left\{2u^\mu\,\nabla_\mu\xi^\al\,n_\al-
(u^\mu \,n^\al+u^\al\, n^\mu) \nabla_\mu\xi_\al\right\}\,\sqrt{\si} d^2 x=\cr
&= {1\over 2\ka}\int_{B_t}
\left\{-2u^\mu\,\xi_\al\,\nabla_\mu n^\al-u^\mu \,n^\al\,\Lie_\xi
g_{\mu\al}\right\}\,\sqrt{\si} d^2 x\cr
 }
\fn
$$
By means of  the identity 
$n^\al\,\Lie_\xi g_{\mu\al}=\Lie_\xi n_\mu-g_{\mu\al}\,\Lie_\xi n^\al$ we obtain:
$$
Q_{\Sigma_t}[\xi,g]={1\over 2\ka}\int_{B_t}\!\!\!
\left\{-2u^\mu\,\xi_\al\nabla_\mu n^\al-u^\mu
(\xi^\al\nabla_\al n_\mu+\nabla_\mu\xi^\al n_\al)
+u_\al\Lie_\xi n^\al\right\}\sqrt{\si}d^2 x
$$
Taking formula $(A.9)$ repeatedly into account together with 
 the condition of orthogonal boundaries  $\left. u^\mu
n_\mu\right\vert_{\calB}=0$ and the condition $\left. \xi^\mu n_\mu\right\vert_{\calB}=0$ we finally
obtain
$$
Q_{\Sigma_t}[\xi,g]={1\over \ka}\int_{B_t}\!\!\!
\left\{\Theta_{\mu\al}\,u^\mu\,\xi^\al\right\}\,\sqrt{\si} d^2 x+
{1\over 2\ka}\int_{B_t}\!\!\!
u_\al\Lie_\xi n^\al\,\sqrt{\si} d^2 x
\fl{\trepiuunoK}
$$
A similar expression may be found for the third contribution $Q_{\Sigma_t}[\xi,\bar g]$
into formula ${\treQ}$, i.e. the Komar contribution of the matched bakground $\bar g$:
$$
Q_{\Sigma_t}[\xi,\bar g]=-{1\over \ka}\int_{B_t}\!\!\!
\left\{\bar \Theta_{\mu\al}\,u^\mu\,\xi^\al\right\}\,\sqrt{\si} d^2 x-
{1\over 2\ka}\int_{B_t}\!\!\!
u_\al{\Lie}_\xi \bar n^\al\,\sqrt{\si} d^2 x
\fl{\trepiuunoKback}
$$
It now remains to calculate the second contribution $Q_{\Sigma_t}[\xi,g, \bar g]$ into
formula ${\treQ}$. We stress that this  is the contribution arising from the boundary
term into the action functional $\FOL$. It can be written as:
$$
\eqalign{
Q_{\Sigma_t}[\xi,g, \bar g]&={1\over 2\ka}\int_{B_t}\!\!\!
(u_\be n_\al-u_\al n_\be)\xi^\al\,w^\be_{\mu\nu}\,g^{\mu\nu}\,\sqrt{\si} d^2 x=\cr
&=-{1\over 2\ka}\int_{B_t}\!\!\!\,u_\al\,\xi^\al\,n_\be\,w^\be_{\mu\nu}\,g^{\mu\nu}\,\sqrt{\si}
d^2 x\cr }
\fn
$$
We remind that in the radial ADM decomposition 
$(A.19)$ of the metric we have $n_\be\,dx^\be=V\, dr$ where $V$ is the radial lapse.
Hence, by making use of the expression $(A.23)$ 
in the Appendix A we obtain:
$$
Q_{\Sigma_t}[\xi,g, \bar g]=-{1\over \ka}\int_{B_t}\!\!\!
 u_\al\,\xi^\al\,(\Theta-\bar \Theta)\,\sqrt{\si} d^2 x
\fl{\trepiuunocorr}
$$
[Notice  that only the projection $u_\al\,\xi^\al$ of the vector field $\xi^\al$ along the
timelike normal $u^\al$ gives a contribution to the term $Q_{\Sigma_t}[\xi,g, \bar g]$.
This is the reason why the boundary term into the action $\FOL$ allows to correct the
anomalous factor of  the Komar superpotential which, as we said above, appears
  in the computation
of mass while it does not enter in  the computation of angular momentum.] 

The 
{\it conserved quantity} in the region  $\Sigma_t$  relative to the infinitesimal
generator of spacetime symmetries $\xi$ is given by the sum of $\trepiuunoK$,
$\trepiuunoKback$ and $\trepiuunocorr$:
$$\eqalign{
Q^{\hbox{\fiverm Tot}}_{\Sigma_t}[\xi,\si]=&{1\over \ka}\int_{B_t}\!\!\!
\left\{\Theta_{\mu\al} \,u^\mu\,\xi^\al-\Theta\,u_\mu\,\xi^\mu \right\}\,\sqrt{\si} d^2 x+\cr
&-{1\over \ka}\int_{B_t}\!\!\!
\left\{\bar\Theta_{\mu\al} \,u^\mu\,\xi^\al-\bar\Theta\,u_\mu\,\xi^\mu
\right\}\,\sqrt{\si} d^2 x+\cr
&+{1\over 2\ka}\int_{B_t}\!\!\!
u_\al\left\{\Lie_\xi n^\al-{\Lie}_\xi \bar n^\al\right\}\,\sqrt{\si} d^2 x\cr
}\fl{\totalQ}
 $$
This latter  formula may be recasted in a form which is better suited to be analysed. Because of  
$\calK_{\mu\nu}u^\mu=0$ and $\si_{\mu}^\nu u^\mu=0$, from formula $(A.11)$ in the Appendix A
it follows in fact:
$$
\Theta_{\mu\al}
\,u^\mu\,\xi^\al=-u_\al\,\xi^\al\,\,n_\mu\,a^\mu-\si^\rho_\al\,\xi^\al\,K_{\rho\be}\,n^\be
\fn
$$
Inserting this latter expression together with $(A.12)$
into $\totalQ$ we obtain:
$$\eqalign{
Q^{\hbox{\fiverm Tot}}_{\Sigma_t}[\xi,\si]=&
-{1\over \ka}\int_{B_t}\!\!\!
\left\{u_\al\,\xi^\al(\calK-\bar{\calK})+
\si^\rho_\al\,\xi^\al\,n^\be(K_{\rho\be}-\bar K_{\rho\be})\right\}\,\sqrt{\si} d^2 x+\cr 
&+{1\over 2\ka}\int_{B_t}\!\!\!
u_\al(\Lie_\xi\,n^\al-{\Lie}_\xi\,\bar n^\al)\,\sqrt{\si} d^2 x\cr
}\fl{\QTOTALE}
$$
Let us stress that, until now, no  assumption has been made on the vector field $\xi$,
apart from the requirement $\left. \xi^\mu n_\mu\right\vert_{\calB}=0$. 

An easy computation shows that {\it the difference}
$u_\al\,(\Lie_\xi\,n^\al-\,{\Lie}_\xi\,\bar n^\al)$ is always zero if the metrics $g$ and $\bar 
g$ are matched on $\calB$.

On the contrary the terms
$u_\al\,\Lie_\xi\,n^\al$ and $u_\al\,\bar {\Lie_\xi}\,n^\al$ appearing in the last contribution to 
$\QTOTALE$  separately  disappear if we choose $\xi$ to be tangent to the $2$--surfaces
$B_t$, that is
$\left.\xi^\mu u_\mu\right\vert_{B_t}=0$, or also if we choose $\xi$ to be 
the time--like  vector field $\partial_t$, i.e. $\xi^\al=N u^\al+N^\al$.
In both cases  the flow of the  vector field
$\xi$ maps each hypersurface $\Sigma_t$ into itself or, respectively, into another surface
$\Sigma_{t'}$. Since the vector field $n^\al$ is tangent to each  $\Sigma_t$ it turns out that also
$\Lie_\xi n^\al$  is tangent to 
 $\Sigma_t$ and then $u_\al\Lie_\xi n^\al=0$ in these cases.

Hence if we specialize formula $\QTOTALE$
for the vector field $\xi=\partial_t$ we obtain the covariant conserved quantity
which we call  the {\it Hamiltonian } of the system. It is given by the expression:
$$
Q^{\hbox{\fiverm Tot}}_{\Sigma_t}[\partial_t,\si]={1\over \ka}
\int_{B_t}\left\{N\,({\cal
K}-\bar {\cal K})-N^a\,(K^{bc}-\bar K^{bc})\, n_b\si_{ac}\right\}\,\sqrt{\si} d^2 x
\fl{\CovADMHam}
$$
and it   coincides exactly with the expression of the $(3+1)$ Hamiltonian $\ADMHamilton$.

Let us notice  that this definition of  Hamiltonian can be correctly considered as the
definition    of a  {\it covariant} ADM formulation (see \ref{\Sinicco}, \ref{\CADM}, 
\ref{\CADMC}). In fact it does not require, {\it a priori}, a ($3+1$) decomposition of spacetime.
We stress that in the covariant ADM approach, the Hamiltonian, or energy, contained in a
$3$--dimensional region
$\Omega$ and relative to a solution $\si$, is defined by $\ConservedQuantity$
as a N\"other conserved quantity:
$$
Q^{\hbox{\fiverm Tot}}_\Omega[\xi,\si]=\int_{\del \Omega} \calU[\xi,\si]
\fl{\CQrich}
$$
This is a  well--posed definition of Hamiltonian provided only that the non--vanishing vector
field
$\xi$be transverse to the hypersurface $\Omega$. Hence, by considering the parameter of the flow
of $\xi$ as the ``time'' parameter and transporting $\Omega$ along the flow of $\xi$ we obtain
a world tube foliated by hypersurfaces diffeomorphic to $\Omega$. In this covariant
context, rather then starting from a preferred local foliation into hypersurfaces, the
starting point is a non--vanishing vector field the  flow of which defines the local time, i.e.
the flow of evolution. Then, by  specializing the definition $\CQrich$ to the ADM foliation
depicted  in  Fig.\ 1, under the additional assumptions of orthogonal boundaries and of the 
matching between the metric and its background, the  {\it covariant} Hamiltonian
$Q^{\hbox{\fiverm Tot}}_{\Sigma_t}[\partial_t,\si]$ exactly coincides with the standard
Hamiltonian  $H(g,\bar g)$ derived from a ($3+1$) splitting of the York action
functional (see equation $\ADMHamilton$).

Another relevant N\"other conserved quantity is obtained by specializing formula
$\QTOTALE$ to a vector field $\xi^\al=N^\al$ tangent to the $2$--surfaces $B_t$,
i.e.
$\left.N^\al u_\al\right\vert_{B_t}=0$ and $\si^\rho_\al\,N^\al=N^\rho$.
 Because of the vanishing of the first and third term in the right hand side of
$\QTOTALE$ we obtain:
$$
Q^{\hbox{\fiverm Tot}}_{\Sigma_t}[N^\al,\si]=
-{1\over \ka}\int_{B_t}\!\!\!
N^\al\,n^\be(K_{\al\be}-\bar K_{\al\be})\,\sqrt{\si}d^2 x
\fl{\ANgMome}
$$
In asymptotically flat spacetimes when $\xi$ corresponds to a  rotation at spatial
infinity,
the N\"other charge $\ANgMome$ may be taken as the definition af {\it angular momentum}.

The last N\"other charge we consider is the one relative to the unit vector field
$\xi=u$ normal to the leaves of the ADM foliation. From $\QTOTALE$  we obtain:
$$
Q^{\hbox{\fiverm Tot}}_{\Sigma_t}[u,\si]=
{1\over \ka}\int_{B_t}\!\!\!
(\calK-\bar{\calK})\,\sqrt{\si}d^2 x
\fl{\Qdiu}
$$
We observe that it agrees with the definition $\qletotale$ of quasilocal energy, i.e. it is the value
of the Hamiltonian $\CovADMHam$ with
$\left.N\right\vert_{B_t}=1$ and
$\left.N^a\right\vert_{B_t}=0$. In the aforementioned hypotheses, quasilocal energy may then be
considered as a N\"other charge.

\NewSection{Time Conservation}

We shall here discuss two different sets of sufficient conditions for time--conservation.
The quantities  $\CovADMHam$, $\ANgMome$ and $\Qdiu$,  are
all covariantly conserved quantities independently on the hypothesis that $\xi$ is a
Killing vector field or not.
In fact they have been defined by means of N\"other theorem
through a construction which relies only on the covariant nature of the Lagrangian.
Hence, on a solution of field equations, the covariant conservation law
$d_\mu\calE^\mu [\xi,\si]=0$ always holds for the N\"other current $\calE[\xi,\si]$ and for all
vector fields $\xi$. This property, together with the property of existence of
superpotentials for any natural theory, has allowed us to define the covariantly
conserved N\"other charges
$Q^{\hbox{\fiverm Tot}}_\Omega[\xi,\si]$ (see
$\ConservedQuantity$). On the contrary the charges $Q^{\hbox{\fiverm Tot}}_{\Sigma_t}[\xi,\si]$ are
{\it conserved in ``time''} if they do not depend on the chosen hypersurface $\Sigma_t$, i.e. if
$Q^{\hbox{\fiverm Tot}}_{\Sigma_t}[\xi,\si]=Q^{\hbox{\fiverm Tot}}_{\Sigma_{t'}}[\xi,\si]$. This is a
stronger condition that has to be supported by additional requirements.  If
$\calB$ is  the $3$--dimensional region such that 
$\Sigma_{t'}-\Sigma_t+\calB$ is the  boundary of a region $D$, from the conservation law 
$d_\mu\calE^\mu[\xi,\si]=0$ we obtain a time--conserved quantity if $\int_\calB
\calE^\mu ds_\mu=0$, i.e. if the net flow of the N\"other current $\calE$ through the hypersurface
$\calB$ vanishes. A stronger condition  amounts to require  the integrand to be  equal to
zero on $\calB$, i.e.
$\left.\calE^\mu\, n_\mu\right \vert_{\calB}=0$. In this case $Q^{\hbox{\fiverm Tot}}[\xi,\si]$ is
conserved not only with respect to the given foliation in hypersurfaces $\Sigma_t$ but it is
time--conserved  with respect to the time of {\it any} foliation of the region $D$.

For the action functional $\FOL$ the N\"other current $\NotherCurrent$ may be rewritten as:
$$
\calE^\al[\xi,\si]={1\over 2\ka}\left\{
(\sqrt{g}\,g^{\mu\nu}-\sqrt{\bar g}\,\bar g^{\mu\nu})\Lie_\xi \bar
u^\al_{\mu\nu}-\Lie_\xi(\sqrt{g}\,g^{\mu\nu})\,w^\al_{\mu\nu}-\xi^\al\,\calL\right\}
\fl{\Erichiamata}
$$
The corresponding quantity  $\left.\calE^\mu\, n_\mu\right \vert_{\calB}$ will be   equal to zero
if  some condition is  verified. We do not explicitly  known a set of necessary  requirements for
the occurence of this situation; we can only provide  two examples of sufficient conditions.

We may require, as a first  example, that the following  properties hold true:
\ss
\itemitem {\bf A)} {the vector field
$\xi$ is  a Killing vector field for the metric, i.e.\ $\Lie_\xi g_{\mu\nu} =0$;}
\itemitem{\bf B)}  {the vector field
$\xi$ is a symmetry for the
background in the sense that $\Lie_\xi\bar u^\al_{\mu\nu}=0$;} 
\itemitem{\bf C)}{
$\xi$ is tangent
to the boundary $\calB$, i.e. $\left.\xi^\mu\, n_\mu\right \vert_{\calB}=0$.} 
\ss
These three requirements together ensure that $\left.\calE^\mu\, n_\mu\right \vert_{\calB}=0$.
These properties  are clearly satisfied  if we are dealing with the Killing vector fields of an
asymptotically flat stationary solution and we choose  the flat metric as a background.
Nevertheless, we stress that  the time--conserved quantities
$Q^{\hbox{\fiverm Tot}}_{\Sigma_t}[\xi,\si]=\int_{B_t}\calU[\xi,\si]$
can be computed on a finite
region, i.e.  it is not necessary that
$B_t$ is identified with  spatial infinity. Moreover, we do not have explicitly required here the
matching between the metrics on 
$\calB$ (see the example of the Kerr metric in Appendix C).

Another set of sufficient conditions may be
imposed to fulfill the condition $\left.\calE^\mu\, n_\mu\right \vert_{\calB}=0$. They closely  resemble the
ones of \ref{\BY}. We may require the matching of the metrics on the boundary $\calB$ 
(so that  the
first term in
$\Erichiamata$ vanishes) and we again require $\xi$ to be tangent
to the boundary $\calB$: $\left.\xi^\mu\, n_\mu\right \vert_{\calB}=0$ (in order to make   the
third term in
$\Erichiamata$ vanishing when contracted with the normal $n_\al$).
We are left with the term:
$$
\int_\calB \calE^\al\,ds_\al=-{1\over
2\ka}\int_\calB \Lie_\xi(\sqrt{g}\,g^{\mu\nu})\,w^\al_{\mu\nu}\,\,\ds_\al
$$
Owing to the matching requirement $\left.g_{\mu\nu}\right \vert_{\calB}=\left.\bar g_{\mu\nu}\right
\vert_{\calB}$ the latter expression may be recasted into the equivalent form:
$$
\int_\calB \calE^\al\,ds_\al=\int_\calB (\Pi^{ij}-\bar \Pi^{ij})\Lie_\xi\ga_{ij}\,d^3 x
$$
It vanishes if we require $\xi$  to be a Killing vector of the boundary $3$--metric:
$\Lie_\xi\ga_{ij}= {\cal D}_i\xi_j+{\cal D}_i\xi_j=0$. Hence, also in this latter situation, we
obtain time--conserved quantities $Q^{\hbox{\fiverm Tot}}_{\Sigma_t}[\xi,\si]$, for the time $t$ of
any foliation of $D$ in hypersurfaces
$\Sigma_t$.

\NewSection{Conclusion and Perspectives}

We proved that once suitable  matching conditions are required
(i.e.\ the $4$--metrics $g$ and $\bar g$ are required to agree on the boundary of the region under
consideration) the two action functionals $\FOL$ and $\YAF$ agree.
Consequently the action functional $\YAF$ may be considered as the ADM counterpart of the
covariant action functional $\FOL$.

The second important result achieved here is the characterization of the quasilocal energy as 
the N\"other charge associated to the (timelike) unit vector normal to the
leaves of the ADM foliation.

These seem to be new results which should enable us to extend
the analysis further ahead to the prescription for the entropy in General Relativity.
In fact, the quasilocal energy as well as the action functional $\YAF$ appeared also as the starting
point of a statistically--oriented approach to black hole entropy (see
\ref{\BYdue }, \ref{\BTZRefB}, \ref{\TaubNUTH},  \ref{\HawHor}, \ref{\TaubNUTHH},
\ref{\TaubNUTHHP} and references quoted therein). 
A different approach to black hole entropy based on N\"other approach (see 
\ref{\BTZ}, \ref{\Remarks}, \ref{\TaubBolt}, \ref{\WaldA} and references quoted therein) may be
found in literature. In view of the present comparison between the covariant first order
approach (which the N\"other approach is based on) and the York's action functional (which
quasilocal energy is based on) as well as between the N\"other charges and the quasilocal energy
themselves, we believe that the two different approaches to entropy can be now successfully
compared (see also
\ref{\Waldqc}, \ref{\Brown}).

Another interesting perpective is to extend the present comparison to the more general situation of
non--orthogonal boundaries which sometimes appeared in the literature (see \ref{\Haywardnon},
\ref{\Mannultimo}, \ref{\HawHun}). It would be of some interest to know whether non--orthogonality
of the boundaries in the ADM decomposition of the covariant action functional $\FOL$ exactly
produces the additional boundary terms which are derived for the (modified) York's action (see
\ref{\Mannultimo}), as we guess to be true.

\bs\ni{\SecTitle Acknowledgments}

This work has been partially supported by INdAM--GNFM and by the 
University of Torino (Italy).
\bs


\NewSection{Appendix A: Notation}
\edef\CurrentSection{A}

In order to make this paper self--contained 
we here briefly summarize some of the formulae and expressions which are used throughout
the paper. We follow the convention and notation  adopted in \ref{\BY} and  we also
refer  the reader to Fig. $1$ for notation.

We assume  the hypersurfaces  $\Sigma_t$ to  be spacelike  and  we assume the
hypersurface  $\calB$ to be timelike. The metric
$h_{\mu\nu}$ induced on the hypersurfaces
$\Sigma_t$ may be written as:
$$h_{\mu\nu}=g_{\mu\nu}+u_\mu\,u_\nu
\fl{\accadig}$$
while for the metric $\gamma_{\mu\nu}$ on the
hypersurface
$\calB$ we have:
$$\gamma_{\mu\nu}=g_{\mu\nu}-n_\mu\,n_\nu
\fl{\gammadig}$$
(in the sequel Greek indices are always raised and lowered with the $4$-- dimensional metric).
The two vectors $\vec u$ and $\vec n$ denote the future directed unit normal to 
$\Sigma_t$ and the outward pointing unit normal to the hypersurfaces 
$\calB$, respectively. They satisfy the normalization relations
$u^\mu u_\mu=-1$ and
$n^\mu n_\mu=1$, respectively.  Any spacetime tensor may be projected onto the
hypersurfaces
$\Sigma_t$ by means of the projection tensor:
$$
h^\mu_\nu=\delta^\mu_\nu+u^\mu u_\nu
\fl{\projh}
$$
Any tensorial object  may  be also projected onto $\calB$ with the projection
tensor
$$
\ga^\mu_\nu=\delta^\mu_\nu-n^\mu n_\nu
\fl{\projga}
$$
The $2$--metric $\sigma_{\mu\nu}$ on
the boundaries
$B_t$ is given by:
$$
\sigma_{\mu\nu}=\ga_{\mu\nu}+u_\mu\,u_\nu=h_{\mu\nu}-n_\mu\,n_\nu
\fl{\sigmadig}
$$
and the respective projection tensor is $\si^\mu_\nu=g^{\mu\rho}\, \si_{\rho\nu}$.

The extrinsic curvatures $K_{\mu\nu}$ of $\Sigma_t$ in $M$, $\Theta_{\mu\nu}$ 
of $\calB$ in $M$
and ${\cal K}_{\mu\nu}$
of
 $B_t$ embedded in $\Sigma_t$ are defined, respectively, as follows:
$$\eqalign{
K_{\mu\nu}=-h^\al_\mu\,\nabla_\al\,u_\nu\cr
\Theta_{\mu\nu}=-\gamma^\al_\mu\,\nabla_\al\,n_\nu\cr
{\cal K}_{\mu\nu}=-\si^\al_\mu\,D_\al\,n_\nu\cr
}\fl{\curvatextr}
$$
where $D_\al$ denotes the covariant derivative on $\Sigma_t$
compatible with the metric $h$.

The extrinsic curvature $K_{\mu\nu}$ is a {\it symmetric tensor on  $\Sigma_t$}, i.e.
 it satisfies the conditions $K_{\mu\nu} u^\mu=0$,   $K_{\mu\nu}
h^\mu_\rho=K_{\rho\nu}$. Instead the extrinsic curvature $\Theta_{\mu\nu}$ is a
symmetric tensor on  $\calB$:
$\Theta_{\mu\nu} n^\mu=0$,  
$\Theta_{\mu\nu} \ga^\mu_\rho=\Theta_{\rho\nu}$, while the extrinsic
curvature ${\cal K}_{\mu\nu}$ is a symmetric tensor on  $B_t$, i.e. ${\cal
K}_{\mu\nu}u^\mu=  {\cal K}_{\mu\nu} n^\mu=0$.

 We also denote by:
$$\eqalign{
a^\nu=u^\mu\nabla_\mu\, u^\nu\cr
b^\nu=n^\mu\nabla_\mu\, n^\nu\cr
}\fl{\quadriacc}
$$
the (covariant) {\it accelerations} of the two normals  $u^\mu$ and $n^\mu$, respectively. They
satisfy the orthogonality properties: $u^\mu\,a_\mu=0$ and $n^\mu\,b_\mu=0$.

By making use of the property
$\projga$,  
we obtain:
$$\nabla_\nu n^\mu=\delta^\al_\nu\,\nabla_\al n^\mu=\gamma^\al_\nu\,\nabla_\al
n^\mu+n^\al\,n_\nu\nabla_\al n^\mu
\fn
$$
Taking into account definitions $\curvatextr$ and $\quadriacc$ we  have:
$$\nabla_\nu n^\mu=-\Theta^\mu_\nu+n_\nu\,b^\mu
\fl{\nablan}$$
Performing calculations in the same manner for the vector $u^\mu$, it is  easy to
check the analogous relation:  
$$\nabla_\nu u^\mu=-K^\mu_\nu-u_\nu\,a^\mu
\fl{\nablau}$$
Moreover, projecting the indices of the extrinsic curvature $\Theta_{\mu\nu}$ normally
and tangentially to the hypersurfaces $\Sigma_t$ we obtain the useful formula (see
\ref{\BY} for detailed computations):
$$\Theta_{\mu\nu}={\cal K}_{\mu\nu}+ u_\mu\,u_\nu\,n_\al a^\al +2 \si_{(\mu}^\al
u^{\phantom{\al}}_{\nu)}\,n^\be\,K_{\al\be}
\fl{\ThetadiK}$$
[We remind that this relation is true only under the assumption  of orthogonal
boundaries, i.e.
$\left.u^\mu\,n_\mu\right\vert_\calB=0$.] Contracting the latter expression with
$g^{\mu\nu}$ we also easily obtain:
$$\Theta={\cal K}-n_\al a^\al
\fl{\tracciaTheta}$$
\medskip\noindent
Let us now consider the ADM decomposition   of the metric:
$$
\eqalign{
g=g_{\mu\nu}\d x^\mu\otimes\d x^\nu=&(h_{\mu\nu}-u_\mu u_\nu)\d x^\mu\otimes\d x^\nu=
                                                                                \cr
=& -N^2\>\d t^2 + h_{ab}(\d x^a+ N^a\>\d t)\otimes (\d x^b+ N^b\>\d t)\cr
}
\fn
$$
The  coordinate system $(x^\mu)=(t,x^a)=(t,r,x^A)$ is adapted  both to $\calB$ and to the
ADM foliation on $D$.
The surfaces $\Sigma_t$ are surfaces of constant $t$ while $\calB$ is the hypersurface
of constant $r$. In other words, indices
$a,b,c,\cdots$ run from $1$ to
$3$ and denote indices on the spacelike hypersurfaces $\Sigma_t$, while 
indices $A,B,C,\cdots$ run from $2$ to
$3$ and they instead denote indices on the boundary $B_t$ of $\Sigma_t$.
Hence, tensors on $\Sigma_t$ are labelled by early Roman letters $a,b,\dots$. 
When they are considered as tensors on spacetime $M$ the same tensors  are instead
denoted by Greek letters. For example, the extrinsic curvature $K$ in $\curvatextr$
can be denoted as $K_{\mu\nu}$ or $K_{ab}$,  according to notational convenience.

 The unit normal
is given by
$ (u_\mu)= (-N, 0,0,0)$ while the timelike coordinate vector field $\vec \partial_0= 
{\partial/\partial t}$ reads as  $\vec \partial_0=N \vec u+ \vec N$, being $N$ the lapse
function and $\vec N= (N^\mu)=(0,N^a)$
the spatial shift vector: $\vec N\cdot \vec u=N^\mu\,u_\mu=0$ (we remind that if we also 
assume the vector field $\vec \partial_0$ be tangent to the $3$--dimensional 
hypersurface
$\calB$, the orthogonal boundaries condition reads as $\left.\vec N\cdot \vec
n\right\vert_\calB=\left.N^\mu\,n_\mu\right\vert_\calB=0$. Nevertheless this latter
hypothesis  will be  not relevant for the computations in the rest of this Appendix). The
extrinsic curvature
$K_{\mu\nu}$ and the acceleration $a^\mu$ defined  in  $\curvatextr$ and
$\quadriacc$, respectively, read as:
$$
\eqalign{
&K_{ab}=-N\,\Gamma^{0}_{ab}={1\over 2N}\,\left[-\partial_0 h_{ab}+D_a N_b+D_b
N_a\right]\cr &(a^\mu)=(0,a^b)=(0,{\partial^b_\cdot N\over N})\cr
}
\fn
$$
(Roman indices are here raised and lowered with the $3$--metric $h_{ab}$).

The Levi--Civita connection coefficients are given by:
$$\eqalign{
&\Gamma^{0}_{0a}={1\over N}(\partial_a N-N^b\,K_{ba})\cr
&\Gamma^{0}_{ab}=-{K_{ab}\over N}\cr
&\Gamma^{0}_{00}={\partial_0 N\over N}+{N^b\over N}\partial_b N-{N^a N^b\over
N}\,K_{ab}\cr 
&\Gamma^{a}_{b0}=-N K^a_b+{N^a N^c\over N}\,K_{cb}+D_b N^a-{N^a\over N}\partial_b
N\cr 
&\Gamma^{a}_{bc}={}^3 \Gamma^{a}_{bc}+{N^a\over N} K_{bc}\cr
}
\fl{\LeviCivsplit}
$$
(where ${}^3 \Gamma^{a}_{bc}$ denotes the Levi--Civita connection of the $3$--metric
$h_{ab}$). A similar splitting may be performed with the background metric $\bar g$; we
obviously obtain the same relations by replacing the objects involved with the
corresponding barred ones. By means of formulae $\LeviCivsplit$ one may easily compute the
quantities $ u^\mu_{\al\be}=  \Ga^{\mu}_{\al\be}-\de^\mu_{(\al}
\Ga^{\ep}_{\be)\ep}$. For example, one can verify the following:
$$g^{\mu\nu}u^0_{\mu\nu}=-{2\over N}  K^a_a+{1\over N^2}\partial_a N^a
\fl{\gperuO}
$$
and 
$$\eqalign{
 g^{\mu\nu}\bar u^0_{\mu\nu}=\bar K_{ab}\left[{1\over N^2\bar N}(\bar N^a-N^a)(\bar
N^b-N^b)-{1\over N}\left({\bar N\over N} \bar h^{ab}+{N\over \bar N} h^{ab}\right)\right]+\cr
-{1\over \bar N N^2}(\bar N^a-N^a)\partial_a\bar N+ {1\over N^2}\partial_a \bar N^a+
{\bar
N^b\over N^2}{}^3\bar\Gamma^a_{ab}-{N^b\over N^2}{}^3\bar\Gamma^a_{ab}
}\fl{\gperbaruO}
$$
which together give the expression 
$$ g^{\mu\nu}w^\al_{\mu\nu} u_\al=
g^{\mu\nu}(u^\al_{\mu\nu}-\bar u^\al_{\mu\nu})u_\al=-N\,g^{\mu\nu}(u^0_{\mu\nu}-\bar
u^0_{\mu\nu})
\fl{\bhouno}
$$
 The latter  term  is involved in  the contribution on the lids of the boundary term in
the action functional $\FOL$. To evaluate instead the action functional contribution on
the  hypersurface
$\calB$ we have to make use of the adapted splitting   of the metric:
$$
\eqalign{
g=g_{\mu\nu}\d x^\mu\otimes\d x^\nu=&(\ga_{\mu\nu}+n_\mu n_\nu)\d x^\mu\otimes\d x^\nu=\cr
=& V^2\>\d r^2 + \ga_{ij}(\d x^i+ V^i\>\d r)\otimes (\d x^j+ V^j\>\d r)\cr
}
\fn$$
where $(x^i)=(t,x^A)$. Middle Roman 
letters  $i,j,k,\cdots$  denote indices on the timelike hypersurface $\calB$ while $x^A$
($A=2,3$) denote again the coordinates over the two  dimensional surfaces $B_t$. The
function $V$ is the {\it radial lapse} while $V^i$ is the {\it radial shift}. Hence the
unit, outward pointing,  radial normal $\vec n$ reads as  $\vec n=(1/V)[\vec
\partial_r -V^i\vec \partial_i]$.
The extrinsic curvature $\Theta_{ij}$ of the ``cylinder'' $\calB=\{r \hbox{\it constant}\}$ is
given by:
$$
\Theta_{ij}=V\,\Gamma^{r}_{ij}={1\over 2V}\,\left[-\partial_r \ga_{ij}+{\cal D}_i
V_j+{\cal D}_j V_i\right]
\fn
$$
where ${\cal D}_i$ denotes the covariant derivative on $\calB$ induced by the
Levi--Civita connection ${}^3 \Ga^i_{jk}$ of the $3$--metric $\ga_{ij}$. 
The  coefficients of the $4$--dimensional Levi--Civita connection can now be decomposed
as:
$$\eqalign{
&\Gamma^{r}_{ri}={1\over V}(\partial_i V+V^j\,\Theta_{ji})\cr
&\Gamma^{r}_{ij}={\Theta_{ij}\over V}\cr
&\Gamma^{r}_{rr}={1\over V}\partial_r V+{V^i\over V}\partial_i V+{V^i
V^j\over V}\,\Theta_{ij}\cr 
&\Gamma^{i}_{jr}=-V \Theta^i_j-{V^i V^k\over V}\,\Theta_{kj}+{\cal
D}_j  V^i-{V^i\over V}\partial_j V\cr 
&\Gamma^{i}_{jk}={}^3 \Gamma^{i}_{jk}-{V^i\over V} \Theta_{jk}\cr
}
\fl{\LeviCivsplitdue}
$$
The latter expressions are not simply obtained from $\LeviCivsplit$ by exchanging tensors on
$\Sigma_t$ with the corresponding tensors on $\calB$.
Because the metric $h_{ab}$ and $\ga_{ij}$ have different signatures, a change of sign
may appear in some terms of $\LeviCivsplitdue$ if compared with the decomposition
$\LeviCivsplit$.
 
By means of $\LeviCivsplitdue$, the following expression are then easily
computed:
$$
\eqalign{
&g^{\mu\nu}u^r_{\mu\nu}={2\over V}  \Theta^i_i-{1\over V^2}\partial_i V^i\cr
&g^{\mu\nu}\bar u^r_{\mu\nu}=\bar \Theta_{ij}\left[{1\over V^2\bar V}(\bar V^i-V^i)(\bar
V^j-V^j)+{1\over V}\left({\bar V\over V} \bar
\ga^{ij}+{V\over \bar V}
\ga^{ij}\right)\right]+\cr 
&\phantom{g^{\mu\nu}\bar u^r_{\mu\nu}}
+{1\over \bar V V^2}(\bar V^i-V^i)\partial_i\bar V- {1\over
V^2}\partial_i \bar V^i- {\bar
V^j\over V^2}{}^3\bar\Gamma^i_{ij}+{V^j\over V^2}{}^3\bar\Gamma^i_{ij}\cr
}\fl{\gperbarur}
$$
Expressions $\gperbarur$  give the expression 
$$
 g^{\mu\nu}w^\al_{\mu\nu}
n_\al=g^{\mu\nu}( u^\al_{\mu\nu}-\bar u^\al_{\mu\nu})n_\al=V\,g^{\mu\nu}( u^r_{\mu\nu}-\bar
u^r_{\mu\nu})
\fl{\bhodue}
$$ 
which is  the contribution on
$\calB$ of the boundary term in the action functional $\FOL$. 

We stress  that in
$\bhouno$ and $\bhodue$ computations are performed without any hypothesis of orthogonal
boundaries and  without requiring any  matching conditions between the metric $g$ and its
background $\bar g$.

\NewSection{Appendix B: Bundle Formalism for Variational Calculus}
\edef\CurrentSection{B}

We hereafter briefly recall the bundle framework for variational calculus and
summarize how one can algorithmically construct the N\"other conserved currents and the
superpotentials out of the Lagrangian.

Let $\calC=(C,M,\pi;F)$ be a bundle and $(x^\mu;y^i)$ be fibered coordinates (relative to a
trivialization chosen on $\calC$).
In fibered coordinates the projection reads as $\pi:(x^\mu;y^i)\mapsto x^\mu$.
A local section defined on $U\subset M$ is a map $\si:U\arr C$ such that
$\pi\circ \si=\one_U$, i.e.\ it is locally given by $\si:x^\mu \mapsto (x^\mu; y^i(x))$.
A local section is thence associated to a map $\si^i:x^\mu\mapsto y^i(x)$.
When $U=M$ the section is called a {\it global section} on $\calC$.
In variational calculus $y^i(x)$ are identified with the values of dynamical fields
at the point $x\in M$. The bundle $\calC$ is called the {\it configuration bundle}.

Starting from the bundle $\pi:C\arr M$ we can define the bundle $\pi^k:J^kC\arr M$
which is called the {\it $k$--order jet prolongation} of $\calC$.
A point in $J^kC$ is denoted by $j^k_x\si$ and it is an equivalence class
of local sections (defined in a neighbourhood of $x\in M$)
having contact of order $k$ at $x$, i.e. having the same Taylor expansion at $x$ up to order $k$.
Points in $J^kC$ are then parametrized by the value of a section $\si$
and its partial derivatives at the point $x\in M$ up to order $k$ included.
Consequently, if $(x^\mu; y^i)$ are fibered ``coordinates'' on $\calC$,
$(x^\mu;y^i, y^i_\mu,\dots, y^i_{\mu_1\dots\mu_k})$ are fibered coordinates on $J^kC$.
We remark that the ``coordinates'' $y^i_{\mu_1\dots\mu_h}$ ($h\le k$) are symmetric in the lower
indices. If $\si$ is a section of $\calC$, we can prolong it to a section $j^k\si$ of $J^kC$
which is defined by $j^k\si(x)= j^k_x\si$.
If $\si:x^\mu\mapsto (x^\mu; y^i(x))$, the $k$--order prolongation $j^k\si$ is given by
$$
j^k\si:x^\mu\mapsto (x^\mu; y^i(x), \del_\mu y^i(x), \dots, \del_{\mu_1\dots \mu_k} y^i(x))
\fn$$
If a section $\si$ is identified with a field configuration, $j^k\si$ describes 
fields and their derivatives up to order $k$. 

Let us denote by $\Phi:C\arr C'$ a bundle morphism projecting onto a diffeomorphism $\phi:M\arr
M'$. By definition of bundle morphism it preserves the fibers (i.e.\ $\pi'\circ \Phi= \phi\circ
\pi$) and it is locally given by
$$
\left\{
\eqalign{
&x'=\phi(x) \cr
&y'=\Phi(x,y)\cr
}
\right.
\fn$$
We can prolong it to a bundle morphism $j^k\Phi:J^kC\arr J^kC'$ defined by the following
(composition preserving) rule
$$
j^k\Phi:j^k_x\si\mapsto j^k_{\phi(x)}\left[\Phi\circ \si\circ \phi^{-1}\right]
\fn$$
(see \ref{\Kolar}, \ref{\Saunder} for further details).
Analogously, for any infinitesimal generator of automorphisms, i.e.\ any projectable vector field 
$\Xi=\xi^\mu(x)\>\del_\mu+\xi^i(x,y)\>\del_i$ on $\calC$,
we can define the prolongation $j^k\Xi=\xi^\mu\>\del_\mu+\xi^i\>\del_i+\xi^i_{\mu}\>\del_i^{\mu}+
\dots+\xi^i_{\mu_1\dots \mu_k}\>\del_i^{\mu_1\dots \mu_k}$ 
(here we set $\del_\mu={\del\over \del x^\mu}$, $\del_i={\del\over \del y^i}$, $\dots$,
$\del_i^{\mu_1\dots \mu_k}={\del\qquad\over \del y^i_{\mu_1\dots \mu_k}}$ for the natural basis
induced in the tangent space $TJ^kC$ by fibered coordinates). It is  recursively given by
$$
\left\{
\eqalign{
&\xi^i_\mu= \d_\mu \xi^i - y^i_{\nu}\del_\mu\xi^\nu
=\d_\mu(\xi^i-y^i_\nu\xi^\nu)+y^i_{\mu\nu}\xi^\nu\cr
&\dots\cr
&\xi^i_{\mu_1\mu_2\dots\mu_k}= \d_{\mu_1} \xi^i_{\mu_2\dots\mu_k} -
y^i_{\nu\mu_2\dots\mu_k}\del_{\mu_1}\xi^\nu
=\d_{\mu_1\dots\mu_k}(\xi^i-y^i_\nu\xi^\nu)+y^i_{\mu_1\dots\mu_k\nu}\xi^\nu
\cr }
\right.
\fn$$
where $\d_\mu$ denotes the {\it total derivative} at any order and it is (locally) defined  by
$$
\d_\mu=\del_\mu + y^i_\mu \>\del_i + y^i_{\nu\mu} \>\del_i^{\nu} +\dots
\fn$$

We define a {\it variational principle of order $k$} to be a pair $(\calC, L)$
where $\calC=(C,M,\pi;F)$ is a bundle called the {\it configuration bundle} and
$L:J^k\calC\arr A_n(M)$ is a (vertical) bundle morphism called the {\it Lagrangian (of order $k$)}.
Here and below $A_h(M)$ ($h\le n$) is the bundle of $h$--forms on $M$ and $n=\dim(M)$.
Locally a Lagrangian reads as $L=\calL(x^\mu,y^i,y^i_\mu,\dots,y^i_{\mu_1\dots\mu_k})\>\ds$ 
(where $\ds=\d x^1\land\dots\land\d x^n$); $\calL$ is the {\it Lagrangian density} and  it depends
on the spacetime point $x^\mu$, on fields $y^i$ and on derivatives of fields
$(y^i_\mu,\dots,y^i_{\mu_1\dots\mu_k})$ up to a finite order $k$.
The Lagrangian $L$ associates to any configuration $\si$ an $n$--form $L\circ j^k\si=(\calL\circ
j^k\si)\>\ds$ over the spacetime $M$.

In General Relativity, the configuration bundle is the bundle  $\calC=\Lor(M)$ of Lorentzian metrics
on the spacetime $M$.
It has local fibered coordinates $(x^\mu; g_{\mu\nu})$. The Hilbert Lagrangian
$L=(1/2\ka)\,\sqrt{g} R\>\ds$ is second order, thus it is interpreted as a map $L:J^2\Lor(M)\arr
A_n(M)$. The second order jet bundle $J^2\Lor(M)$ has natural fibered coordinates
$(x^\mu, g_{\mu\nu}, g_{\mu\nu,\la}, g_{\mu\nu, \la\si})$ or, equivalently, it can be
parametrized by the more convenient set of variables
$(x^\mu, g_{\mu\nu}, \Ga^\la_{\mu\nu}, \d_\si\Ga^\la_{\mu\nu})$
where $\Ga^\la_{\mu\nu}$ denote the Christoffel symbols of the metric $g$.

We stress that in this framework, the Lagrangian is a map between finite--dimensional manifolds.
It induces the action functional:
$$
A_D(\si)=\int_D L\circ j^k\si
\fn$$
where $D\subset M$ is a region in  spacetime $M$, i.e.\ a compact submanifold (of dimension $n$)
with a boundary $\del D$ which is again a compact submanifold (of dimension $n-1$).

Let now $\calC=(C,M,\pi;F)$ be a fiber bundle. 
We can define the sub--bundle $\tau_{_{V}}:V(C)\arr C$ of the tangent bundle $\tau:TC\arr C$,
which turns out to be a vector sub--bundle of $TC$. A vector $v\in T_pC$ belongs to $V_pC$ if and
only if
$T_p\pi(v)=0$; in this case we say that $v$ is a {\it vertical vector at $p\in C$}.
Accordingly, $V(C)$ is defined to be the kernel $\ker(T\pi)$ of the fiberwise linear map
$T\pi:TC\arr TM$. A vertical vector is ``tangent to the fiber'' $F$ and its local expression
(using fibered coordinates $(x^\mu; y^i)$ on $\calC$) is $v=v^i\>\del_i$ where $\del_i={\del\over
\del y^i}$ is a local pointwise basis of vertical vectors.
A section of the bundle $V(C)\arr C$ is called a {\it vertical vector field}.
When $\calC$ is the configuration bundle of a field theory, a vertical vector field
$X=(\de y^i)\>\del_i$ represents a {\it deformation} of the dynamical fields.
We can also consider the dual vector bundle $V^\ast(C)$ of pointwise linear functionals on
vertical vectors, which is called the {\it dual vertical bundle}.
A pointwise basis $\bar{\d} y^i$ of $V^\ast(C)$ can be defined to be the dual basis to $\del_i$,
i.e.\ $\bar{\d} y^i(\del_j)=\de^i_j$.
Notice that, even if $V(B)\subset T(C)$, $V^\ast(C)$ is not a sub--bundle of $T^\ast(C)$.

We stress that the dual vertical bundle can be defined for any bundle; we can consider the dual
vertical bundle $V^\ast(C)$ of the configuration bundle as well as the vector bundle 
$V^\ast(J^kC)$ of its prolongations. The importance of these objects is that they allow to {\it
translate} the variation of the action functional into the variation of the
Lagrangian morphism.
For any Lagrangian $L:J^k\calC\arr A_n(M)$ we can in fact define a bundle morphism
$\de L:J^{k}C\arr V^\ast(J^kC)\otimes A_n(M)$ such that for all vertical vector fields $X:C\arr V(C)$
we have:
$$
<\de L\>\vert\> j^kX>= \left({\d\over \d s} (L\circ j^k\Phi_s)\right)_{s=0}
\fn$$
where $\Phi_s:C\arr C$ is the (vertical) flow of $X$ and $<\>\vert\>>$ denotes the canonical duality
between $V^\ast(J^kC)$ and $V(J^kC)$ given by evaluation.
Locally, the morphism $\de L$ is given by
$$
\de L= \left[ p_i\>\bar{\d} y^i + p_i^{\mu}\>\bar{\d} y^i_{\mu}+ \dots +
 p_i^{\mu_1\dots\mu_k}\>\bar{\d} y^i_{\mu_1\dots\mu_k}\right]\otimes \ds
\fn$$
where we defined the {\it conjugate momenta} $(
p_i=\del_i\calL,\>p_i^{\mu}=\del_i^{\mu}\calL,\>\dots,
\>p_i^{\mu_1\dots\mu_k}=\del_i^{\mu_1\dots\mu_k}\calL)$
to be the derivatives of the Lagrangian density
$\calL$ with respect to the ``coordinates''
$(y^i,y^i_\mu,\dots,y^i_{\mu_1\dots\mu_k})$, respectively.

The variation of the action along a deformation $X\in V(C)$ is thence given by
$$
\de_X A_D(\si)=\int_D <\de L\>\vert \> j^kX>\circ j^k\si
\fl{\blach}
$$
Field equations now follow by Hamilton principle, i.e.\ requiring the action to be stationary
along any deformation around classical solutions.
In other words, we say that a configuration $\si:M\arr C$ is a {\it classical solution}
if, for all regions $D$ and for all deformations $X$ such that $j^{k-1}X\vert_{\del D}=0$
(i.e.\ such that the deformation $X$ vanishes together with its derivatives up to order $k-1$ on
the boundary $\del D$ of the region $D$) we have:
$$
\de_X A_D(\si)=0 
\fn$$ 
Specializing to the Lagrangian $\FOL$, which is first order in $g$ and second order in $\bar g$, the
deformation $X=X_g+X_{\bar g}= (\de g^{\mu\nu})\del_{\mu\nu}+(\de \bar g^{\mu\nu})\bar\del_{\mu\nu}$
splits into a deformation $X_g$ of the dynamical metric $g^{\mu\nu}$ which is required to vanish  on
the boundary $\del D$ (i.e.\ $X_g\vert_{\del D}=0$) and a deformation $X_{\bar g}$ of the background
metric $\bar g^{\mu\nu}$ which is required to vanish together with its first order derivatives
(i.e.\
$j^1X_{\bar g}\vert_{\del D}=0$) on the boundary $\del D$.

One can prove that there exist a (unique) global bundle morphism $\E:J^{2k}C\arr V^\ast(C)\otimes
A_n(M)$ (called the {\it Euler--Lagrange morphism}) and (a family of) global bundle morphisms
$\F:J^{2k-1}C\arr V^\ast(J^{k-1}C)\otimes A_{n-1}(M)$ (called the {\it Poincar\'e--Cartan
morphisms}) such that the so--called {\it first variation formula} holds true, i.e.:
$$
<\de L\>\vert\> j^k X> \circ j^k\si= <\E\>\vert\> X>\circ j^{2k}\si + \d \left(<\F \>\vert\> j^{k-1}
X>\circ j^{2k-1}\si\right)
\fl{\FVF}$$
Since in the Hamilton principle we consider deformations such that
$j^{k-1}X\vert_{\del D}=0$ the Poincar\'e--Cartan contribution vanishes when integrated in
$\blach$ and we get field equations $\E\circ j^{2k}\si=0$, i.e. Euler--Lagrange field equations.

If, for example, we consider $k=2$ we obtain
$$
\left\{
\eqalign{
& \E=[p_i - \d_\mu p_i^\mu + \d_{\mu\nu} p_i^{\mu\nu}]\bar d y^i\otimes \ds\cr
& \F=[(p_i^\mu-\d_\nu p_j^{\mu\nu})\>\bar{\d} y^i
+p_i^{\mu\nu}\>\bar{\d} y^i_\nu]\otimes\ds_\mu\cr
 }
\right.
\fn$$

A bundle $\calC=(C,M,\pi;F)$ is called {\it natural} if there exists a natural lift to $C$ of
spacetime diffeomorphisms, i.e.\ if one can associate to any spacetime diffeomorphism $\phi:M\arr
M$ a (unique) bundle automorphism $\Phi:C\arr C$ so to preserve compositions.
As a consequence, on natural bundles one can canonically and naturally lift spacetime vector
fields
$\xi=\xi^\mu(x)\>\del_\mu$ to bundle vector fields $\hat\xi= \xi^\mu(x)\>\del_\mu+
\xi^i(x,y)\>\del_i$. A field theory is called a {\it natural theory} if its configuration bundle
$\calC$ is a natural bundle and, in addition, any spacetime diffeomorphism is a Lagrangian
symmetry, i.e.\ we have
$$
<\de L\>\vert\> j^k \Lie_{\xi}\si> \circ j^k\si= \Lie_\xi (L\circ j^k\si)
\fl{\Covariance}$$
where $\Lie_\xi (L\circ j^k\si)=(\d \circ i_\xi + i_\xi\circ \d) (L\circ j^k\si)$ is the {\it
Lie derivative of the Lagrangian} and $\Lie_{\xi}\si=T\si\circ \xi-\hat\xi\circ\si$ is the {\it Lie
derivative of the section $\si$} along a spacetime vector field $\xi$. Locally we have
$$
\Lie_{\xi}\si= (\xi^\mu\> \d_\mu y^i(x)- \xi^i(x,y(x)))\>\del_i
\fn$$
Notice that $\Lie_{\xi}\si$ is a vertical vector field on the section $\si$
so that the l.h.s.\ of formula $\Covariance$ is meaningful.
We also recall that equation $\Covariance$ is equivalent to the requirement that the so--called
{\it Poincar\'e--Cartan form} is  invariant under  the flow of $\xi$ (see \ref{\Lagrange}).

The global N\"other theorem is now easily obtained.
Let $L$ be a natural Lagrangian; inserting $\FVF$ into eqation $\Covariance$  we obtain
$$
<\E\>\vert\> \Lie_{\xi}\si> +\, \d <\F\>\vert\> j^{k-1}\Lie_{\xi}\si>= \d \>(i_\xi L)
\fn
$$
i.e.
$$
\d(<\F\>\vert\> j^{k-1}\Lie_{\xi}\si> - i_\xi L)=- <\E\>\vert\> \Lie_{\xi}\si>
\fn$$
The {\it N\"other current} and the {\it work density} are thence defined by
$$
\eqalign{
&\calE(L,\xi,\si)= <\F\>\vert\> j^{k-1}\Lie_{\xi}\si> - i_\xi L\cr
&\calW(L,\xi,\si)=-<\E\>\vert\> \Lie_{\xi}\si>\cr
}
\fn$$
and they satisfy
$$
\d\calE(L,\xi,\si)=\calW(L,\xi,\si)
\fl{\Bsedici}
$$
In any natural theory, both the N\"other current $\calE(L,\xi)$ and the work density\
$\calW(L,\xi)$ can be recasted, by suitable covariant integration by parts, as:
$$
\eqalign{
&\calE(L,\xi,\si)= \tilde\calE(L,\xi,\si)+\d(\calU(L,\xi,\si))\cr
&\calW(L,\xi,\si)= \calB(L,\xi,\si)+\d(\hat\calE(L,\xi,\si))\cr
}
\fl{\Bdicia}
$$
One can easily prove that $\calB(L,\xi,\si)=0$ off--shell (i.e.\ on any configuration not
necessarily a solution of field equations), that $\hat\calE(L,\xi,\si)=\tilde\calE(L,\xi,\si)$ and
that $\tilde\calE(L,\xi,\si)$ vanishes on--shell (i.e.\ along solutions of field equations).
The identity $\calB(L,\xi,\si)=0$ is called the {\it generalized Bianchi identity};
the current $\tilde\calE(L,\xi,\si)$ is called the{\it reduced current}, while $\calU(L,\xi,\si)$
is called the {\it superpotential}.
We stress  once again that all the objects introduced are algorithmically constructed out of
the Lagrangian.

We remark (see $\Bsedici$) that on shell the N\"other current $\calE(L,\xi,\si)$ obeys a 
continuity equation $\d \calE(L,\xi,\si)=0$, i.e.\ it is covariantly conserved.
The  conservation laws of this kind are called {\it weak conservation laws} since they hold on
solutions of field equations only, in opposition to the {\it strong conservation laws} which hold
also
off--shell. An example of strongly conserved quantity is $\calE(L,\xi,\si)-\tilde\calE(L,\xi,\si)$
which is a closed form along any configuration $\si$, even if it is not a solution of
field equations (see $\Bdicia$).

An example of second order natural theory is General Relativity. Some details about this case are
found in Section $5$. Further details can be found in \ref{\Lagrange}, \ref{\Robutti},
\ref{\Remarks}. The N\"other currents, as long as the superpotentials, are used in Section $5$
to define covariantly conserved quantities in the particular case of General Relativity.

\NewSection{Appendix C: Examples}
\edef\CurrentSection{C}

We shall here discuss some simple examples to illustrate and clarify some of the topics introduced above in the
paper. 
\medskip
Let us first consider the conserved quantities of the Schwarzschild solution given in its
standard form:
$$
g=-\left(1- {2M\over \rho}\right)\>\dt^2 + {1\over \left(1- {2M\over \rho}\right)}
 \d \rho^2 
+\rho^2\>(\d \te^2+\sin^2\te\>\d \phi^2)
\fn
$$
For computational convenience we rewrite it in isotropic coordinates $(t,r,\te,\phi)$:
$$
g=-{(2r-M)^2\over (2r+M)^2}\>\dt^2 + \left(1+{M\over 2r}\right)^4\Big( \d r^2 
+r^2\>(\d \te^2+\sin^2\te\>\d \phi^2)\Big)
\fl{\schisotrope}$$
where, see \ref{\Gravitation}:
$$
2r= -M+\rho+\sqrt{\rho(\rho-2M)}\qquad \Longleftrightarrow\qquad \rho={(M+2r)^2\over 4r}
\fl{\rdirho}
$$
Let us choose as a background the Minkowski spacetime in isotropic coordinates
$$
\bar g=-{(2R-M)^2\over (2R+M)^2}\>\dt^2 + \left(1+{M\over 2R}\right)^4\Big( \d r^2 
+r^2\>(\d \te^2+\sin^2\te\>\d \phi^2)\Big)
\fn$$
The background $\bar g$ is in fact a flat metric as one can easily check by direct computation of
the Riemann tensor. Furthermore, the two metrics $g$ and $\bar g$ are matched on the hypersurface
$\calB$ defined by the equation  $r=R$.

Let us then consider a family of foliations of the region $r\le R$ generated by the infinitesimal
generator
$$
\ze=\big(1+\ep\>(\sin(\phi)+1)\big)\>\del_t
\fn$$
where $\ep$ is a (small) parameter.
The flow parameter is denoted by $s$ so that, if $\Sigma$ is a spacelike  hypersurface, it can be
dragged along  the flow $\Phi_s$ of $\ze$ to obtain the foliation:
$$
\Si_s^\ep=\{\hbox{$s$ {\it constant}}\} \qquad s=t/(1+\ep\>(\sin(\phi)+1))
$$
Since  $s$ denotes the affine parameter along the flow of $\ze$, it can be interpreted as 
 the {\it time} associated
to the ADM foliation. We remark that we are in the hypothesis of orthogonal boundaries as required throughout
the paper. Notice also that for $\ep=0$ we recover the ordinary asymptotic time translation
$\ze=\del_t$ and the ordinary ADM foliation by the hypersurfaces $t=${\it constant} (see Fig.\ $2$).

{
\hskip30truept 
\TuboDritto 
\vskip-117pt
\hskip160pt
 \TuboStorto

\hskip70truept $\ep=0$ \hskip100pt $\ep\not=0$
\vskip 8pt   \hskip70truept  Fig. $2$A \hskip90pt Fig. $2$B

\vskip-147pt  \hskip 10pt $\Si_{s=1}$ \hskip205pt $\Si^\ep_{s=1}$
\vskip 30pt   \hskip 10pt $\Si_{s=0}$ \hskip205pt $\Si^\ep_{s=0}$
\vskip 38pt   \hskip 10pt $\Si_{s=-1}$ \hskip200pt $\Si^\ep_{s=-1}$

\vskip 63truept
}

Let us then choose the ($3$--parameter) vector field 
$$\xi=\al\>\del_t+ (\be+\ga\>\sin(\te)\>\cos\phi)\del_\phi
\fn
$$
(where $\al$, $\be$ and $\ga$ are three real constants)
which is a generator of symmetries for the first order Lagrangian
$\FOL$. We remark that $\xi$ is a well defined vector field on $\calB$; in particular it extends to
$\te=0$ and $\te=\pi$. For $\al=1$ and $\be=\ga=0$, $\xi$ reduces to the ordinary  time
translation
$\xi=\del_t$ so that we expect the corresponding conserved quantity $Q^{\hbox{\fiverm
Tot}}_{\Si^\ep_s}[\xi]$ to be interpreted as
the {\it mass} of $g$ {\it relative} to $\bar g$ on the leaf $\Si^\ep_s$ in the region $r\le R$.
We also stress that for $\ga=0$ the vector field $\xi$ is a Killing vector both for $g$
and $\bar g$.
On the contrary, for $\ga\not= 0$, $\xi$ is not Killing both  for $g$ and $\bar g$.  In the case 
$\ga\not= 0$, $\xi$ has not a direct physical interpretation though 
 it is a symmetry generator for the Lagrangian $\FOL$ and it algorithmically generates the  covariant N\"other 
conserved quantity $Q^{\hbox{\fiverm Tot}}_{\Si^\ep_s}[\xi]$.
We shall use it to illustrate how time conservation may be related to covariant conservation
along different foliations.

If we calculate the N\"other conserved quantity (according to eq.\ $\qsigmat$  with $2k=16\pi$, i.e.
in geometric units $G=c=1$) we get the following result
$$
\eqalign{
&Q_{\Si_s^\ep}[\xi, g]= \al{M\over 2}- \ga \>{\pi\over 16} \> R \left(1-{M\over 2R}\right)^2
s\>\ep\cr
&Q_{\Si_s^\ep}[\xi, g,\bar g]=
 \left(\al{M\over 2}-\ga\>{\pi M\over 16}\> s\>  \ep\right)\left(1-{M\over R}\right)\cr
&Q_{\Si_s^\ep}[\xi, \bar g]=
 \ga\>{\pi R\over 16}\left(1-{M^2\over 4 R^2}\right)\> s\>\ep
\cr
&Q^{\hbox{\fiverm Tot}}_{\Si_s^\ep}[\xi]=\al \left( M -{M^2\over 2R}\right)
+\ga\left({\pi M^2\over 32 R}s\> \ep \right)\cr
}
\fl{\AppR}
$$

Thus, when $\xi$ is the Killing vector $\del_t$ ($\al=1$, $\be=\ga=0$) the {\it relative mass}
$$
Q^{\hbox{\fiverm Tot}}_{\Si_s^\ep}[\del_t]= M -{M^2\over 2R}
\fl{\massarelativa}
$$
is time--conserved along any foliation of the family
generated by $\ze$, since it does not depend on the affine parameter $s$.
We also stress that the {\it conserved quantity} 
(i.e.\ letting $R$ tend to infinity)
is always $Q^{\hbox{\fiverm Tot}}_\infty[\del_t]=  M $, i.e.\ it reduces to the expected value for
total mass.

If $\ga\not=0$ the vector $\xi$ is not Killing. Despite of the fact that the conserved
quantity is not time--conserved in general, it is still time--conserved along a particular foliation
(namely, the one  corresponding to $\ep=0$, i.e.\ $\ze=\del_t$; see Fig.\ $2$A).
We remark that in this case the flow of the N\"other current $\calE[\xi]=\calE^\mu\,\ds_\mu$
through $\del \Sigma_s$ vanishes
even if
$\left. \calE^\mu\,n_\mu\right\vert_{\calB}\neq 0$.

On the contrary, when $\ep\not=0$ (see Fig.\ $2$B) 
the N\"other charge $Q^{\hbox{\fiverm Tot}}_{\Si_s^\ep}$ explicitly
depends on the time
$s$, i.e.\ on the particular leaf on which it is computed.
 
We finally remark that in any case the quantity $Q^{\hbox{\fiverm Tot}}_{\Si_s^\ep}[\xi]$ given by
equation $\AppR$ is covariantly conserved (as any N\"other conserved quantity is) and it
consequently obeys a continuity equation.

To end up this  first example let us now compute the quasilocal energy $\Qdiu$. We consider the case depicted
in Fig. $2$A, i.e. the foliation induced by $\ze=\partial_t$. We choose, as infinitesimal symmetry
$\xi$, the  unit timelike normal  $u$:
$$
\xi=u={2r+M\over 2r-M}\partial_t
\fn
$$
If we calculate the N\"other conserved quantity \ $\qsigmat$  
we now obtain  the following result:
$$
\eqalign{
&Q_{\Si_t}[u, g]={1\over 4} {M(2R+ M)\over 2R-M}\cr
&Q_{\Si_t}[u, g,\bar g]= {1\over 2}{(2R^2-MR-M^2)M\over R(2R-M)}\cr
&Q_{\Si_t}[u, \bar g]= {1\over 4}{M(4R^2-M^2)\over (2R-M)^2}\cr
&Q^{\hbox{\fiverm Tot}}_{\Si_t}[u]=M +{M^2\over 2R} \cr
}
\fl{\locenrsch}$$
In  spherical coordinates $(t,\rho,\theta, \phi)$ the fourth  expression of
$\locenrsch$ may be rewritten as:
$$
Q^{\hbox{\fiverm Tot}}_{\Si_t}[u]=\rho_0\left[  1- \sqrt{1-{2M\over \rho_0}}\right]
\fl{\qleBYas} 
$$
where $2R= -M+\rho_0+\sqrt{\rho_0(\rho_0-2M)}$; see
$\rdirho$.
As expected,  expression $\qleBYas$ perfectly agrees with the value of the  energy computed  in
\ref{\BY}, formula
$(6.14)$.

\bigskip

Let us now consider another example of conserved quantity in a finite region $D$.
It is a completely different example since it does not require the match of the solutions
on the boundary $\calB$ of the finite region under consideration. Neither the condition of orthogonal 
boundaries is here required.

 Let us consider the Kerr spacetime in ingoing Kerr--Schild coordinates
$(t,r,\te,\phi)$, given by:
$$
g=\bar g+ 2Mr\rho^{-2}\Big[\dt +\dr-a\sin^2\theta\dphi\Big]^2
\fn$$
where $\rho^2=r^2+a^2\cos^2\theta$, $M^2\ge a^2$. Let us choose  the flat background $\bar g$ as:
$$
\bar g= -\dt^2+\Big[\dr-a\sin^2\theta\>\dphi\Big]^2+\rho^2\Big[\d\theta^2+\sin^2\theta\>\dphi^2\Big]
\fn
$$
The metrics $g$ and $\bar g$ are matched at infinity. Let us consider the regions $D$ {\it inside}
the hypersurface $\calB$ defined by $r=R$ and the ADM foliation $\Si_t=\{t \hbox{\it constant}\}$
generated by the vector field $\del_t$.
We stress that $g$ and $\bar g$ do not match on $\calB$ unless $R$ is let to tend to infinity.

Let us finally choose as symmetry generator the ($2$--parameter) vector
$$
\xi=\al\>\del_t + \be\>\del_\phi
\fn$$
which is a Killing vector for   $g$ and $\bar g$ ($\al$ and $\be$ are two real constants).

The N\"other conserved quantities one obtains are
$$
\eqalign{
&Q_{\Si}[\xi, g]=\left( \al\>\hbox{$M\over 2$} -\be\> Ma \right)\cr
&Q_{\Si}[\xi, g,\bar g]=\al\hbox{$M\over 2$} \cr
&Q_{\Si}[\xi, \bar g]=0 \cr
&Q^{\hbox{\fiverm Tot}}_{\Si}[\xi]=\al\>M -\be\> Ma\cr
}
\fl{\cqkerr}$$
which reproduce the expected values of the {\it relative mass } and of the {\it angular momentum}
in the region $t=$constant and $r\le R$.
Notice that the result is independent on $R$ meaning that all the energy and angular momentun is
``buried in the singularity''.
We remark that setting anywhere $a=0$ the Schwarzschild solution is recovered. The relative mass $M$ we obtain
in this case (see equation $\cqkerr$ with $\al=1$, $\be=0$) does not agree  with the value found  above (see
equation $\massarelativa$) because of the two different matches selected.

We remark that the quantities $\cqkerr$ are also time--conserved (even if the metrics are not
matched at
$\calB$).
In fact the flow integral $\int_\calB \calE$  vanishes since $\xi$ is tangent to $\calB$ and it is
a Killing vector of both $g$ and $\bar g$ (see Section $7$).
Notice also that $\calE^\al\> n_\al=0$, i.e.\ the N\"other current has no flow through any part of
$\calB$.
Consequently, the associated conserved quantity is time conserved along {\it any} ADM foliation of
the region $D$.

[The calculations in this Appendix have been carried out by using tensor package of MapleV,  see
\ref{\McLenaghan}. They are the very direct application of formula $\qsigmat$, just computed on
the configuration $(g,\bar g)$.]
 
\vfill\eject

\NewSection{References}

\Biblio

\end